\documentclass[unsortedaddress,aps,preprint]{revtex4}
\usepackage{graphicx}
\usepackage{dcolumn}
\usepackage{amsmath}
\usepackage{amssymb}
\usepackage{amsfonts}
\usepackage{tikz}
\usepackage{bm}
\usepackage{color}
\usepackage[normalem]{ulem}
\usepackage{hyperref}
\hypersetup{colorlinks=true,linkcolor=blue,urlcolor=magenta}
\newcommand{\be}{\begin{equation}}
\newcommand{\ee}{\end{equation}}
\newcommand{\ba}{\begin{eqnarray}}
\newcommand{\ea}{\end{eqnarray}}
\newcommand{\ve}[1]{\left| #1\right>}
\newcommand{\veb}[1]{\big| #1\big>}
\newcommand{\me}[2]{\left< #1|#2\right>}
\newcommand{\meb}[2]{\big< #1\big|#2\big>}
\newcommand{\va}{\left|\right>}
\newcommand*\circled[1]{\tikz[baseline=(char.base)]{\node[shape=circle,draw,inner sep=2pt] (char) {#1};}}

\begin{document}
\title{Bose-Hubbard Model on Polyhedral Graphs}
\author{Santi Prestipino$^1$\footnote{Email: {\tt sprestipino@unime.it}}}
\affiliation{$^1$Universit\`a degli Studi di Messina,\\Dipartimento di Scienze Matematiche e Informatiche, Scienze Fisiche e Scienze della Terra,\\viale F. Stagno d'Alcontres 31, 98166 Messina, Italy}
\date{\today}

\begin{abstract}
Ever since the first observation of Bose-Einstein condensation in the nineties, ultracold quantum gases have been the subject of intense research, providing a unique tool to understand the behavior of matter governed by the laws of quantum mechanics. Ultracold bosonic atoms loaded in an optical lattice are usually described by the Bose-Hubbard model or a variant of it. In addition to the common insulating and superfluid phases, other phases (like density waves and supersolids) may show up in the presence of a short-range interparticle repulsion and also depending on the geometry of the lattice. We herein explore this possibility, using the graph of a convex polyhedron as ``lattice'' and playing with the coordination of nodes to promote the wanted finite-size ordering. To accomplish the job we employ the method of decoupling approximation, whose efficacy is tested in one case against exact diagonalization. We report zero-temperature results for two Catalan solids, the tetrakis hexahedron and the pentakis dodecahedron, for which a thorough ground-state analysis reveals the existence of insulating ``phases'' with polyhedral order and a widely extended supersolid region. The key to this outcome is the unbalance in coordination between inequivalent nodes of the graph. The predicted phases can be probed in systems of ultracold atoms using programmable holographic optical tweezers.
\end{abstract}
\maketitle

\section{Introduction}

The last few decades have seen a development of very effective atom-cooling methods~\cite{Phillips} that has eventually culminated in the first observation ever of Bose-Einstein condensation in atomic gases~\cite{Anderson,Davis}. Concurrently, also the ability to manipulate laser beams has been continuously increasing, to the point that one can create periodic potentials of various dimensionality (``optical lattices'') which are free of defects and stable~\cite{Windpassinger}. Optical trapping of ultracold atoms provides an invaluable means to probe the behavior of quantum particles on a lattice, thus representing a desirable platform for the study of collective effects in many-body quantum systems~\cite{Jaksch,Greiner,Bloch}.

In the original Bose-Hubbard (BH) model~\cite{Fisher}, the competition between itinerant and localized character of quantum states is reduced to the bone: kinetic energy, represented through a $U(1)$-invariant hopping term, is made minimum by a broken-symmetry condensed state spread over the entire volume of the system, whereas potential energy favors localization of particles. As a result, at zero temperature ($T=0$) the system exists in either a superfluid or an insulating ground state, with a quantum transition between them. The scenario becomes richer when the range of interaction between particles increases. Then, depending on the lattice, other insulating ground states (ordinary solids) may appear; moreover, crystalline order may coexist with superfluidity (supersolids). Earlier examples of supersolid ground states in extended BH models have been reported in \cite{vanOtterlo,Goral,Sengupta,Kovrizhin}, while the first observations of a density-modulated structure coexisting with phase coherence are more recent~\cite{Tanzi,Boettcher,Chomaz}.

We here expand the catalogue of spinless boson systems where density waves, either with off-diagonal long-range order or not, are stable at $T=0$ by considering finite ``lattices'', or better polyhedral graphs (i.e., made up from the vertices and edges of a polyhedron) as underlying supporting frame for the particles. While clearcut phases and phase transitions are not possible on a finite graph, the absence of natural boundaries and a relatively high symmetry in the spatial distribution of nodes make our investigation valuable for a comparison with ordinary lattice models. Our interest goes to regular or semiregular polyhedra inscribed in a sphere, since these ensure sufficient homogeneity in the coordination of vertices, a property shared with lattices. The use of spherical boundary conditions (SBC) has often been exploited in the past to discourage long-range ordering at high density~\cite{Post,Prestipino,Prestipino2,Prestipino3,Vest,Guerra,Franzini,Prestipino4,Prestipino5}. On the other hand, SBC make it possible to observe forms of ordering that are unknown to Euclidean space. An added value of a spherical mesh is the possibility to vary the coordination of vertices while keeping the overall geometry strictly two-dimensional (a polyhedral graph is a planar graph). From the point of view of experiment, we note that bosons confined in thin spherical shells (``bubble traps'') have already been realized~\cite{Zobay,Garraway} and will soon be studied in microgravity~\cite{Elliott,Lundblad}. Present laser-light technology based on optical tweezers already has the sophistication needed to constrain atoms within a close neighborhood of the vertices of a chosen polyhedron~\cite{Barredo,Browaeys}.

A preliminary study of the extended BH model on the graph of a regular polyhedron has been given in Ref.~\cite{Prestipino6}. There, we have employed the decoupling approximation~\cite{Fisher,Sheshadri} (DA, a kind of mean-field theory) to sketch the phase behavior at $T=0$, finding that DA is already reliable for a graph as simple as that of a cube. Here, we carry out a similar analysis for more complex graphs, choosing the skeleton of two Catalan solids for demonstration. As in \cite{Prestipino6} we make the further simplification that multiple node occupancy is forbidden, which corresponds to a system of hard-core bosons. With this assumption, the dimensionality of the Hilbert space is reduced to such a degree that in one case the DA can be validated against exact diagonalization. The main lesson of the present investigation is that, when the vertex set of a graph can be decomposed into a few subsets of inequivalent vertices, then the superfluid phase is ruled out and a wide supersolid region appears in its place. Thus, semiregular graphs are an ideal playground where to observe supersolid ``phases'', in addition to insulating ``solids'' with polyhedral symmetry.

The rest of the paper is organized as follows. In Section 2 we describe the model, the physical observables of interest, and the method used to perform the investigation. There is not a unique way to motivate the DA method, and we have devoted a few appendices to present various equivalent derivations of this approximation for the reader's benefit. Section 3 contains the core of our study: in Sections 3.A to 3.C we illustrate our theory for the graph of a tetrakis hexahedron, which is still sufficiently simple to be amenable to exact analysis. Then, in Section 3.D we focus on the graph of a pentakis dodecahedron and repeat the DA treatment of the extended BH model. Finally, concluding remarks follow in Section 4.

\section{Model and theory}
\setcounter{equation}{0}
\renewcommand{\theequation}{2.\arabic{equation}}

In its simplest terms, the grand Hamiltonian of the extended BH model on a regular lattice reads
\be
H=-t\sum_{\langle i,j\rangle}\big(a_i^\dagger a_j+a_j^\dagger a_i\big)+\frac{U}{2}\sum_in_i(n_i-1)+V\sum_{\langle i,j\rangle}n_in_j-\mu\sum_in_i\,,
\label{2-1}
\ee
where $a_i,a_i^\dagger$ are bosonic field operators and $n_i=a_i^\dagger a_i$ is a number operator. Moreover, $t\ge 0$ is the hopping amplitude between nearest-neighbor (NN) sites, $U>0$ is the on-site repulsion, $V>0$ is the strength of the NN repulsion favoring the spatial distancing of bosons, and $\mu$ is the chemical potential. Were it not for the hopping term, the BH model would not be dissimilar from a classical lattice gas, sharing with it the same sequence of phases as a function of $\mu$. Things change completely with the inclusion of quantum kinetic energy, which makes it possible for particles to be delocalized even at $T=0$, a situation that goes along with a macroscopic occupation of the zero-momentum state. When $V\ne 0$, the interplay between insulating and superfluid order may generate so-called supersolid phases where both crystalline and superfluid order are simultaneously present~\cite{Pollet,Ng,Iskin,Kimura,Ohgoe}. In the hard-core limit $U\rightarrow +\infty$, the site occupancy will be effectively restricted to zero or one and the $U$ term in (\ref{2-1}) can be discarded; following a well-established tradition~\cite{vanOtterlo,Wessel,Kurdestany,Zhang,Yamamoto,Gheeraert}, it is only this limit that is treated hereafter.

%
%
\begin{figure}
\begin{center}
\begin{tabular}{cc}
{\tt Tetrakis Hexahedron}\qquad\qquad & {\tt Pentakis Dodecahedron}\\
\includegraphics[width=7cm]{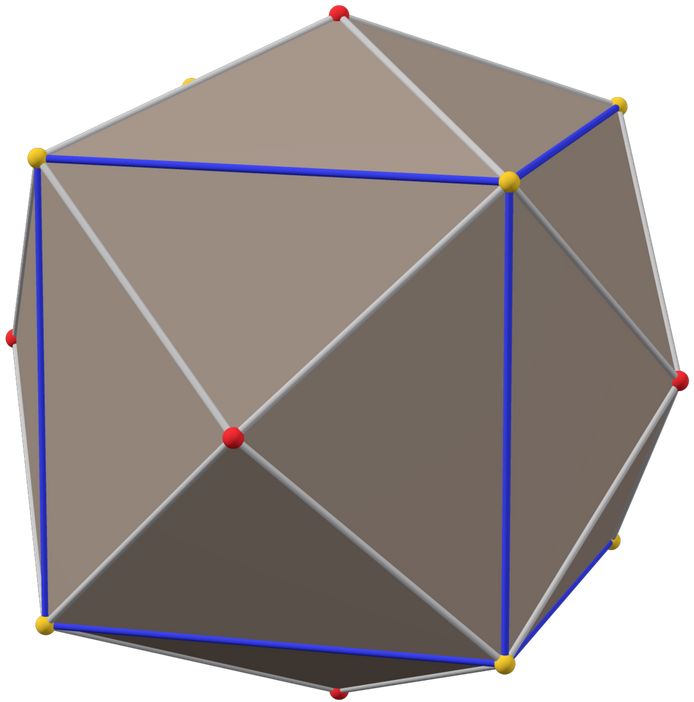}\qquad\qquad & \includegraphics[width=7cm]{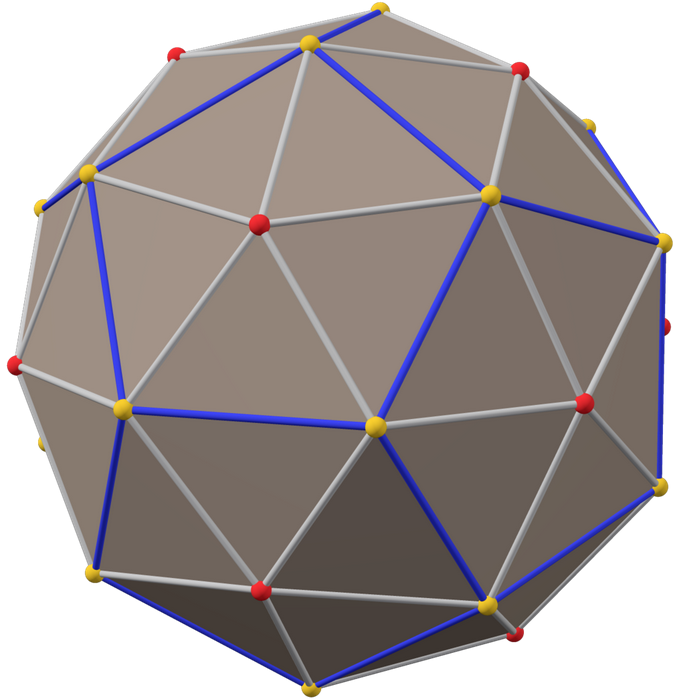}
\end{tabular}
\end{center}
\caption{The two Catalan solids considered in this work. Left: tetrakis hexahedron (TH), with 6 octahedral vertices (red dots) and 8 cubic vertices (yellow dots); each cubic vertex belongs to either of two different tetrahedral subsets. Right: pentakis dodecahedron (PD), with 12 icosahedral vertices (red dots) and 20 dodecahedral vertices (yellow dots). In both pictures, the long edges are colored in blue and the short edges are colored in grey. Two nodes of the graph are considered NN if they are joined by an edge of the polyhedron, either long or short.}
\label{fig1}
\end{figure}

%
%
\begin{table}
\caption{Main elements of the polyhedra considered in the present study. The quoted lengths refer to the biscribed form of the polyhedron and are in units of the circumscribed radius (i.e., all vertices of the biscribed polyhedron lie on the sphere of radius 1)}
\begin{center}
\begin{tabular}{ccc}
\hline\hline
& {\rm tetrakis hexahedron} & {\rm pentakis dodecahedron}\\
\hline
vertices & $14\,\,\,(6\,[4]+8\,[6])$ & $32\,\,\,(12\,[5]+20\,[6])$\\
faces & $24\,\,\,({\rm isosceles\,\,triangles})$ & $60\,\,\,({\rm isosceles\,\,triangles})$\\
edges & $36\,\,\,(24\,\,{\rm short}\,+\,12\,\,{\rm long})$ & $90\,\,\,(60\,\,{\rm short}\,+\,30\,\,{\rm long})$\\
symmetry & ${\rm full\,\,octahedral\,\,(Oh)}$ & ${\rm full\,\,icosahedral\,\,(Ih)}$\\
short edge & $\sqrt{6\big(3-\sqrt{3}\big)}\big/3=0.9194\ldots$ & $\sqrt{30\left(15-\sqrt{15\big(5+2\sqrt{5}\big)}\right)}\big/15=0.6408\ldots$\\
long edge & $2\sqrt{3}/3=1.1547\ldots$ & $\big(\sqrt{15}-\sqrt{3}\big)/3=0.7136\ldots$\\
circumscribed radius & 1 & 1\\
inscribed radius & \,\,\,\,\,\,$1\big/\sqrt{5-2\sqrt{3}}=0.8068\ldots$\,\,\,\,\,\, & \,\,\,\,\,\,$1\big/\sqrt{10-\sqrt{5}-\sqrt{6\big(5+\sqrt{5}\big)}}=0.9226\ldots$\,\,\,\,\,\,\\
volume & $8/3=2.6666\ldots$ & $2\sqrt{10\big(5-\sqrt{5}\big)}\big/3=3.5048\ldots$\\
\hline\hline
\end{tabular}
\end{center}
\end{table}

In Ref.\,\cite{Prestipino6} we have studied model (\ref{2-1}) at $T=0$ on a polyhedral graph with $M$ nodes, focusing on those Platonic polyhedra (i.e., the cube and the dodecahedron) where a subset of vertices forms itself a regular polyhedron. Besides a number of insulating ``phases'', crystalline or not, the ground-state diagram contains a wide superfluid basin and, only in the dodecahedral case, a small supersolid region. In this paper, the hosting space for bosons is still the graph of a convex polyhedron, but now taken to be {\em semiregular}. Our choice goes in particular to Catalan solids, which are isohedral (i.e., all faces are equivalent under the symmetries of the figure) but neither isogonal (vertices are not all equivalent) nor circumscribable. Among this class of polyhedra, the two which are simplicial (have triangular faces) and deviate less from isogonality are the {\em tetrakis hexahedron} (TH, Kleetope of a cube and dual to the truncated octahedron) and the {\em pentakis dodecahedron} (PD, Kleetope of a dodecahedron and dual to the truncated icosahedron), see Fig.\,1. To make them circumscribable, the pyramids added to each face of the cube (TH) or dodecahedron (PD) are adjusted in height so that the solid, already inscribable, becomes also circumscribable --- with this change, the deviation from isogonality is slightly reduced. We collect in Table I the main characteristics of the biscribed forms of TH and PD. We note that $T=0$ cluster ``phases'' with TH and PD symmetry are found in a system of soft-core bosons on the sphere~\cite{Prestipino4}.

Compared to a Platonic solid, each polyhedron in Fig.\,1 has two species of vertices and also two kinds of edges, long and short. Therefore, in view of interpreting the Hamiltonian (\ref{2-1}) clearly, we are faced with the problem of choosing between two notions of nearness on the graph: one possibility is that NN nodes are exclusively those joined by a short edge (then, the ends of a long edge are second-neighbor nodes). On the other hand, we may decide to call NN the pairs of nodes that are adjacent in the graph, namely joined by an edge of the polyhedron, regardless of being long or short. Clearly, the nature of BH phases changes from one case to the other. Free from obligations dictated by phenomenology, we can base our choice on the kind of phase sequence we want at $t=0$. It turns out that the phase diagram is richer if we use adjacency as criterion of nearness, as we do in the following.

Once the hosting graph has been chosen, we analyze the $T=0$ phase diagram of the extended BH model with $U=+\infty$ using the DA. In short, we linearize the hopping and repulsion terms in (\ref{2-1}) using~\cite{Prestipino6}
\be
a_i^\dagger a_j\approx a_i^\dagger\left<a_j\right>+\big<a_i^\dagger\big>a_j-\big<a_i^\dagger\big>\left<a_j\right>\,\,\,\,\,\,{\rm and}\,\,\,\,\,\,n_in_j\approx n_i\left<n_j\right>+\left<n_i\right>n_j-\left<n_i\right>\left<n_j\right>\,,
\label{2-2}
\ee
where the ground-state averages $\left<a_i\right>\equiv\phi_i$ and $\left<n_i\right>\equiv\rho_i$ are to be determined self-consistently. $\phi_i$ and $\rho_i$ represent the superfluid order parameter and local density for the $i$-th site, respectively (the condensed fraction is $|\phi_i|^2$). The simplified Hamiltonian reads
\be
H_{\rm DA}=-t\sum_i\big(F_ia_i^\dagger+F_i^*a_i-F_i\phi_i^*\big)+\frac{V}{2}\sum_i\left(2R_in_i-R_i\rho_i\right)-\mu\sum_in_i
\label{2-3}
\ee
with $F_i=\sum_{j\in{\rm NN}_i}\phi_j$ and $R_i=\sum_{j\in{\rm NN}_i}\rho_j$. We refer the reader to Appendix A to C for a thorough justification of this approximation. The self-consistency equations for the parameters $\phi_i$ and $\rho_i$ are also the conditions under which the grand potential of (\ref{2-3}) is stationary, see Appendix B.

\section{Results}
\setcounter{equation}{0}
\renewcommand{\theequation}{3.\arabic{equation}}

By the DA, the original problem of determining the grand potential of (\ref{2-1}) is reduced to the much simpler task of diagonalizing the one-site Hamiltonian (\ref{2-3}). At $T=0$, only the minimum eigenvalue and its eigenstate are needed. For the graph of a semiregular polyhedron, the job is even simpler since we can identify a few inequivalent subsets of the vertex set and, from the viewpoint of mean-field (MF) theory, assume that the order parameters are homogeneous in each subset (i.e., a single creation operator can be used to populate a whole subset of vertices). In Ref.\,\cite{Prestipino6}, where in the cases investigated the vertex subsets are two, the strategy put forward was to diagonalize a two-site Hamiltonian, hence a $4\times 4$ matrix. Here, we find easier to divide the same task in as many one-site problems as are the vertex types, which are three for both TH and PD graphs.

\subsection{TH model}

Looking at Fig.\,1a, the fourteen TH vertices can be classified as octahedral (6) or cubic (8), implying a natural decomposition of the TH graph into two inequivalent groups of vertices. However, with an interaction that is repulsive at NN separation, we may expect a different number and superfluid density in the two subsets of tetrahedral vertices of which the set of cubic vertices is made up. Hence, we find it necessary to divide the vertices of the TH graph in three subsets, A, B, and C, consisting of the octahedral, tetrahedral-1, and tetrahedral-2 nodes, respectively, and accordingly write the MF Hamiltonian (\ref{2-3}) as a function of six order parameters. Since
\ba
&&F_{\rm A}=2\phi_{\rm B}+2\phi_{\rm C}\,,\,\,\,F_{\rm B}=3\phi_{\rm A}+3\phi_{\rm C}\,,\,\,\,F_{\rm C}=3\phi_{\rm A}+3\phi_{\rm B}\,;
\nonumber \\
&&R_{\rm A}=2\rho_{\rm B}+2\rho_{\rm C}\,,\,\,\,R_{\rm B}=3\rho_{\rm A}+3\rho_{\rm C}\,,\,\,\,R_{\rm C}=3\rho_{\rm A}+3\rho_{\rm B}\,,
\label{3-1}
\ea
the MF Hamiltonian reads:
\small
\ba
H_{\rm DA}&=&E_0-12t\left[(\phi_{\rm B}+\phi_{\rm C})a_{\rm A}^\dagger+(\phi_{\rm B}^*+\phi_{\rm C}^*)a_{\rm A}\right]-12t\left[(\phi_{\rm A}+\phi_{\rm C})a_{\rm B}^\dagger+(\phi_{\rm A}^*+\phi_{\rm C}^*)a_{\rm B}\right]
\nonumber \\
&-&12t\left[(\phi_{\rm A}+\phi_{\rm B})a_{\rm C}^\dagger+(\phi_{\rm A}^*+\phi_{\rm B}^*)a_{\rm C}\right]+6(2V\rho_{\rm B}+2V\rho_{\rm C}-\mu)n_{\rm A}
\nonumber \\
&+&4(3V\rho_{\rm A}+3V\rho_{\rm C}-\mu)n_{\rm B}+4(3V\rho_{\rm A}+3V\rho_{\rm B}-\mu)n_{\rm C}
\label{3-2}
\ea
\normalsize
with
\be
E_0=12t\left[(\phi_{\rm B}+\phi_{\rm C})\phi_{\rm A}^*+(\phi_{\rm A}+\phi_{\rm C})\phi_{\rm B}^*+(\phi_{\rm A}+\phi_{\rm B})\phi_{\rm C}^*\right]-12V\left[\rho_{\rm A}\rho_{\rm B}+\rho_{\rm A}\rho_{\rm C}+\rho_{\rm B}\rho_{\rm C}\right]\,.
\label{3-3}
\ee
For fixed $t$ and $\mu$, the matrix representing the DA Hamiltonian on the canonical basis $\ve{x_{\rm A},x_{\rm B},x_{\rm C}}$ (with $x_i=0$ or 1) is $8\times 8$. The simplest case is $t=0$, where the matrix becomes diagonal. Then, each basis vector is an energy eigenvector and the corresponding diagonal element is the eigenvalue. While $\phi_{\rm A}=\phi_{\rm B}=\phi_{\rm C}=0$, the density parameters are calculated by making each eigenvalue stationary; for the eigenvalue of $\ve{x_{\rm A},x_{\rm B},x_{\rm C}}$ we obtain $\rho_{\rm A}=x_{\rm A},\rho_{\rm B}=x_{\rm B}$, and $\rho_{\rm C}=x_{\rm C}$. With these parameters, the minimum eigenvalue for the given $\mu$ yields the grand potential $\Omega$, and its eigenvector is the ground state. We observe a ``phase transition'' when the relative stability of two eigenvalues changes. Clearly, on a finite graph only a smooth crossover may occur, any thermodynamic singularity being an artifact of MF theory. Results for $t=0$ are summarized in the table below:
\ba
\begin{tabular}{cccc}
$\mu\,\,{\rm range}$\qquad\qquad & {\rm grand\,\,potential}\qquad\qquad & {\rm ground\,\,state}\qquad\qquad & phase\\
\hline
$\mu\le 0$\,:\qquad\qquad & $0$\qquad\qquad & $\ve{0,0,0}$\qquad\qquad & {\rm ``empty''}\\
$0\le\mu\le 3V$\,:\qquad\qquad & $-6\mu$\qquad\qquad & $\ve{1,0,0}$\qquad\qquad & {\rm OCT}\\
$3V\le\mu\le 6V$\,:\qquad\qquad & $12V-10\mu$\qquad\qquad & $\ve{1,1,0}\,{\rm and}\,\ve{1,0,1}$\qquad\qquad & {\rm OCT+TET}\\
$\mu\ge 6V$\,:\qquad\qquad & $36V-14\mu$\qquad\qquad & $\ve{1,1,1}$\qquad\qquad & {\rm ``full''}
\nonumber\\
\end{tabular}
\ea
To be clear, ``empty'' is the phase with no particle at all; ``OCT'' is the phase where all the octahedral nodes are occupied ($N=6$ particles in total); ``OCT+TET'' is the two-fold degenerate phase where either A and B or A and C are filled ($N=10$); finally, ``full'' is the phase with one particle at each node ($N=14$). It is worth noting that, should we have opted for a notion of nearness as proximity in space, we would have got a stable CUB phase (i.e., one with only the cubic nodes occupied) for $0\le\mu\le 4V$, in addition to ``empty'' ($\mu\le 0$) and ``full'' ($\mu\ge 4V$).

For $t>0$, the minimum eigenvalue $\lambda_{\rm min}$ of the Hamiltonian matrix is most easily obtained by separately diagonalizing a $2\times 2$ matrix in each vertex subset (see Appendix A). The equations for $\rho_i$ and $\phi_i$ are then obtained by making $\lambda_{\rm min}$ stationary. It is a simple matter to show that
\small
\ba
\lambda_{\rm min}&=&E_0+12V(\rho_{\rm A}+\rho_{\rm B}+\rho_{\rm C})-7\mu-3\sqrt{(2V\rho_{\rm B}+2V\rho_{\rm C}-\mu)^2+16t^2|\phi_{\rm B}+\phi_{\rm C}|^2}
\nonumber \\
&-&2\sqrt{(3V\rho_{\rm A}+3V\rho_{\rm C}-\mu)^2+36t^2|\phi_{\rm A}+\phi_{\rm C}|^2}-2\sqrt{(3V\rho_{\rm A}+3V\rho_{\rm B}-\mu)^2+36t^2|\phi_{\rm A}+\phi_{\rm B}|^2}\,.
\nonumber \\
\label{3-4}
\ea
\normalsize
For superfluid and supersolid phases, $\phi_{\rm A},\phi_{\rm B}$, and $\phi_{\rm C}$ are generally non-zero complex numbers. However, these parameters should have equal phases since only the magnitude of the order parameter can be spatially modulated. Without loss of generality, we may take the arbitrary phase as zero, implying that $\phi_{\rm A},\phi_{\rm B}$, and $\phi_{\rm C}$ are positive quantities. With this specification, the equations for the parameters are considerably simplified and become the following:
\ba
\rho_{\rm B}+\rho_{\rm C}&=&1-\frac{3V\rho_{\rm A}+3V\rho_{\rm C}-\mu}{2\sqrt{\circled{\rm B}}}-\frac{3V\rho_{\rm A}+3V\rho_{\rm B}-\mu}{2\sqrt{\circled{\rm C}}}\,;
\nonumber \\
\rho_{\rm A}+\rho_{\rm C}&=&1-\frac{2V\rho_{\rm B}+2V\rho_{\rm C}-\mu}{2\sqrt{\circled{\rm A}}}-\frac{3V\rho_{\rm A}+3V\rho_{\rm B}-\mu}{2\sqrt{\circled{\rm C}}}\,;
\nonumber \\
\rho_{\rm A}+\rho_{\rm B}&=&1-\frac{2V\rho_{\rm B}+2V\rho_{\rm C}-\mu}{2\sqrt{\circled{\rm A}}}-\frac{3V\rho_{\rm A}+3V\rho_{\rm C}-\mu}{2\sqrt{\circled{\rm B}}}\,;
\nonumber \\
\phi_{\rm B}+\phi_{\rm C}&=&\frac{3t(\phi_{\rm A}+\phi_{\rm C})}{\sqrt{\circled{\rm B}}}+\frac{3t(\phi_{\rm A}+\phi_{\rm B})}{\sqrt{\circled{\rm C}}}\,;
\nonumber \\
\phi_{\rm A}+\phi_{\rm C}&=&\frac{2t(\phi_{\rm B}+\phi_{\rm C})}{\sqrt{\circled{\rm A}}}+\frac{3t(\phi_{\rm A}+\phi_{\rm B})}{\sqrt{\circled{\rm C}}}\,;
\nonumber \\
\phi_{\rm A}+\phi_{\rm B}&=&\frac{2t(\phi_{\rm B}+\phi_{\rm C})}{\sqrt{\circled{\rm A}}}+\frac{3t(\phi_{\rm A}+\phi_{\rm C})}{\sqrt{\circled{\rm B}}}
\label{3-5}
\ea
with
\ba
\circled{\rm A}&=&(2V\rho_{\rm B}+2V\rho_{\rm C}-\mu)^2+16t^2(\phi_{\rm B}+\phi_{\rm C})^2\,;
\nonumber \\
\circled{\rm B}&=&(3V\rho_{\rm A}+3V\rho_{\rm C}-\mu)^2+36t^2(\phi_{\rm A}+\phi_{\rm C})^2\,;
\nonumber \\
\circled{\rm C}&=&(3V\rho_{\rm A}+3V\rho_{\rm B}-\mu)^2+36t^2(\phi_{\rm A}+\phi_{\rm B})^2\,.
\label{3-6}
\ea
Apparently, the above set of non-linear equations cannot be solved exactly. To overcome the problem, we can numerically minimize a non-negative function $G$ of the order parameters, constructed in such a way as to vanish when Eqs.~(\ref{3-5}) and (\ref{3-6}) are simultaneously fulfilled. For given $t$ and $\mu$ values, we generate a grid of points in parameter space, which is then made finer and finer around each zero of $G$ where $\lambda_{\rm min}$ is low, until the best parameters and the absolute minimum $\Omega$ of (\ref{3-4}) are determined with sufficient precision. Typically, several competing minima may occur, which suggests that one should proceed carefully to avoid that some zero of $G$ may escape the net.

%
%
\begin{figure}
\begin{center}
\includegraphics[width=16cm]{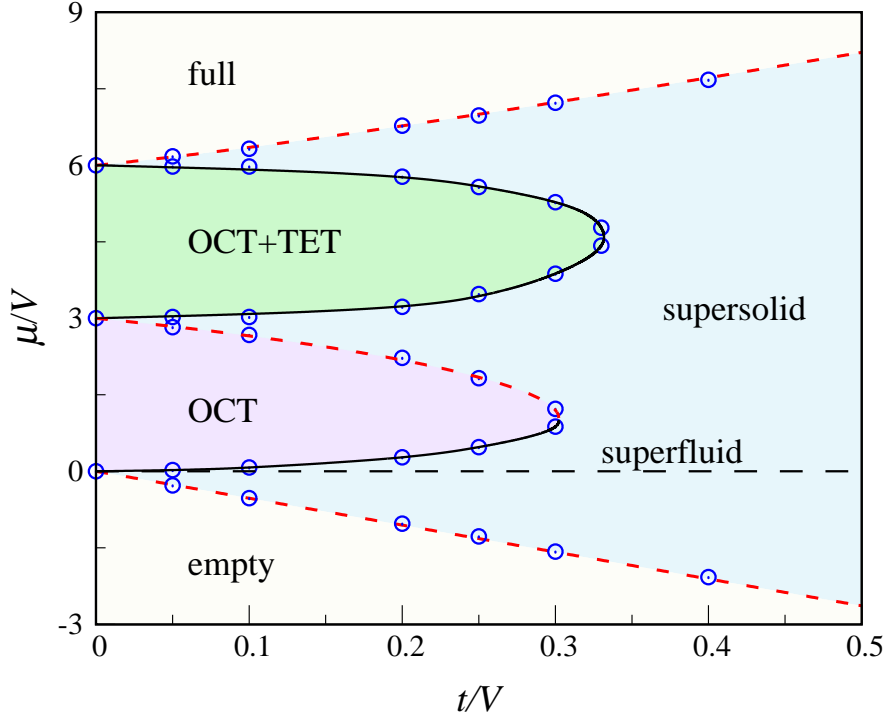}
\end{center}
\caption{MF phase diagram of the extended BH model with $U=+\infty$ on a TH graph, using $V$ as unit of energy. The blue dots mark transition points. The long-dashed $\mu=0$ line is the only place where the system is superfluid. The dashed red curves are the continuous-transition loci derived in the text (cf. Eqs.~(\ref{3-11}), (\ref{3-14}), and (\ref{3-17})). The remaining transitions lines are first-order.}
\label{fig2}
\end{figure}

We sketch in Fig.\,2 the resulting MF phase diagram at $T=0$. The dots are phase-transition points at which the solution to Eqs.~(\ref{3-5}) and (\ref{3-6}) changes qualitatively. As a result, particles can exist in five distinct phases, four insulating and one supersolid (SS). For each $t=0$ phase with polyhedral order, there is a lobe in the $(t,\mu)$ plane where the same order persists up to a certain $t$, before SS eventually prevails. In the latter phase, $\phi_{\rm A}\ne\phi_{\rm B}=\phi_{\rm C}$ and $\rho_{\rm A}\ne\rho_{\rm B}=\rho_{\rm C}$, to within the numerical uncertainty of our computation. A superfluid phase only exists along the line $\mu=0$: if we take $\rho_{\rm A}=\rho_{\rm B}=\rho_{\rm C}=\rho$ and $\phi_{\rm A}=\phi_{\rm B}=\phi_{\rm C}=\phi$ in Eqs.~(\ref{3-5}) and (\ref{3-6}), we readily obtain
\be
\rho=\frac{t}{V+2t}\,\,\,\,\,\,{\rm and}\,\,\,\,\,\,\phi=\frac{\sqrt{Vt+t^2}}{V+2t}\,\,\,\,\,\,\longrightarrow\,\,\,\,\,\,\Omega=-\frac{36t^2}{V+2t}\,.
\label{3-7}
\ee

%
%
\begin{figure}
\begin{center}
\includegraphics[width=8.1cm]{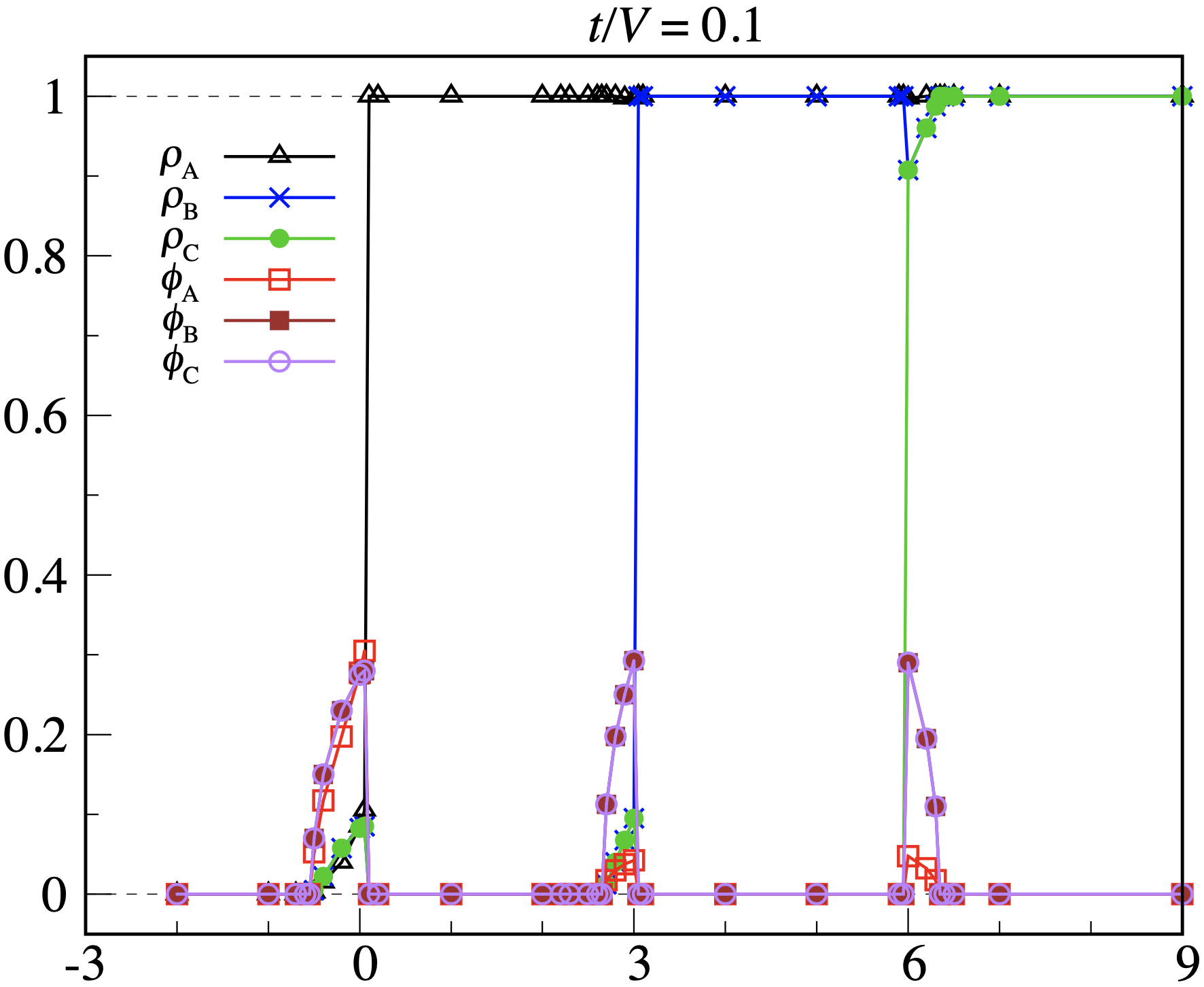}
\includegraphics[width=8.1cm]{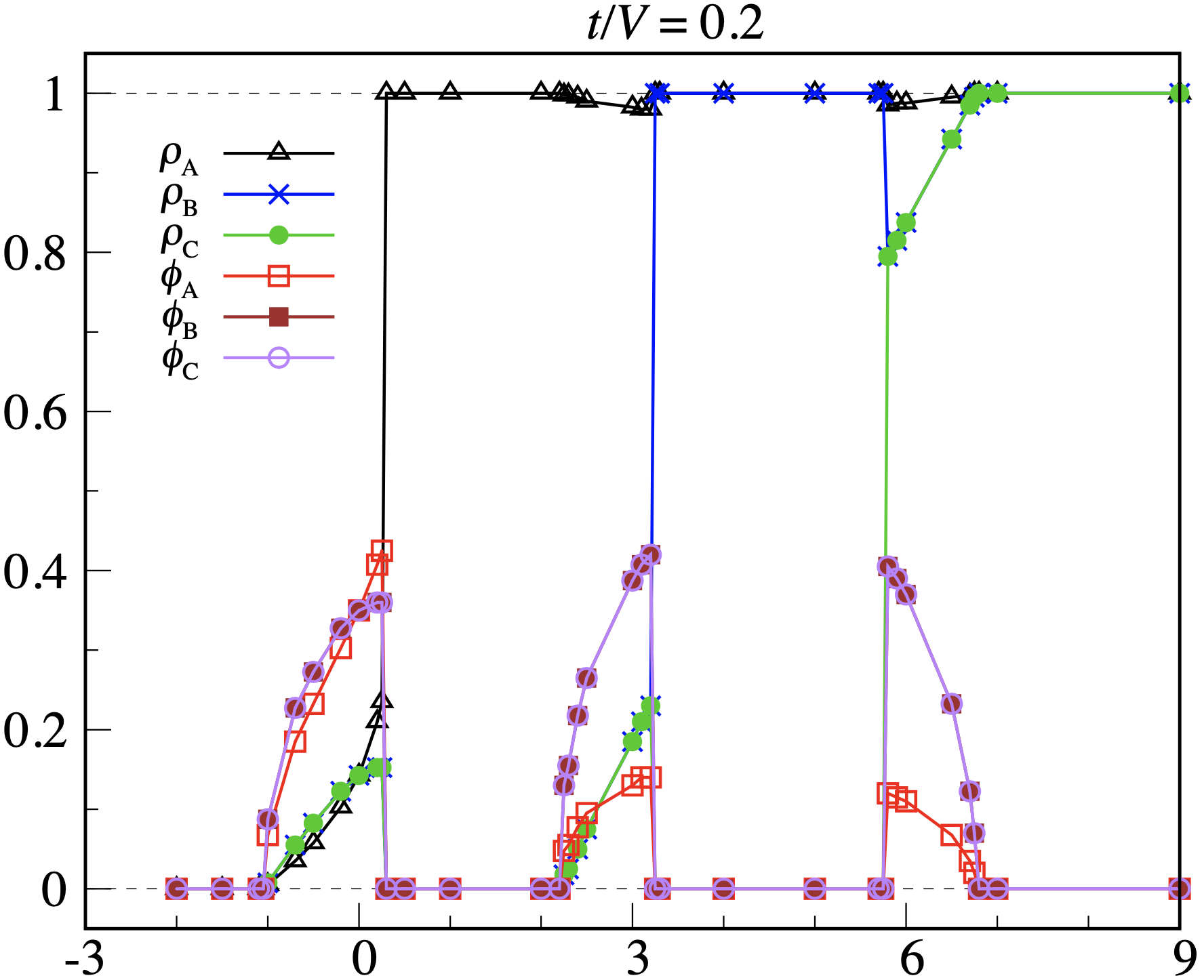}
\includegraphics[width=8.1cm]{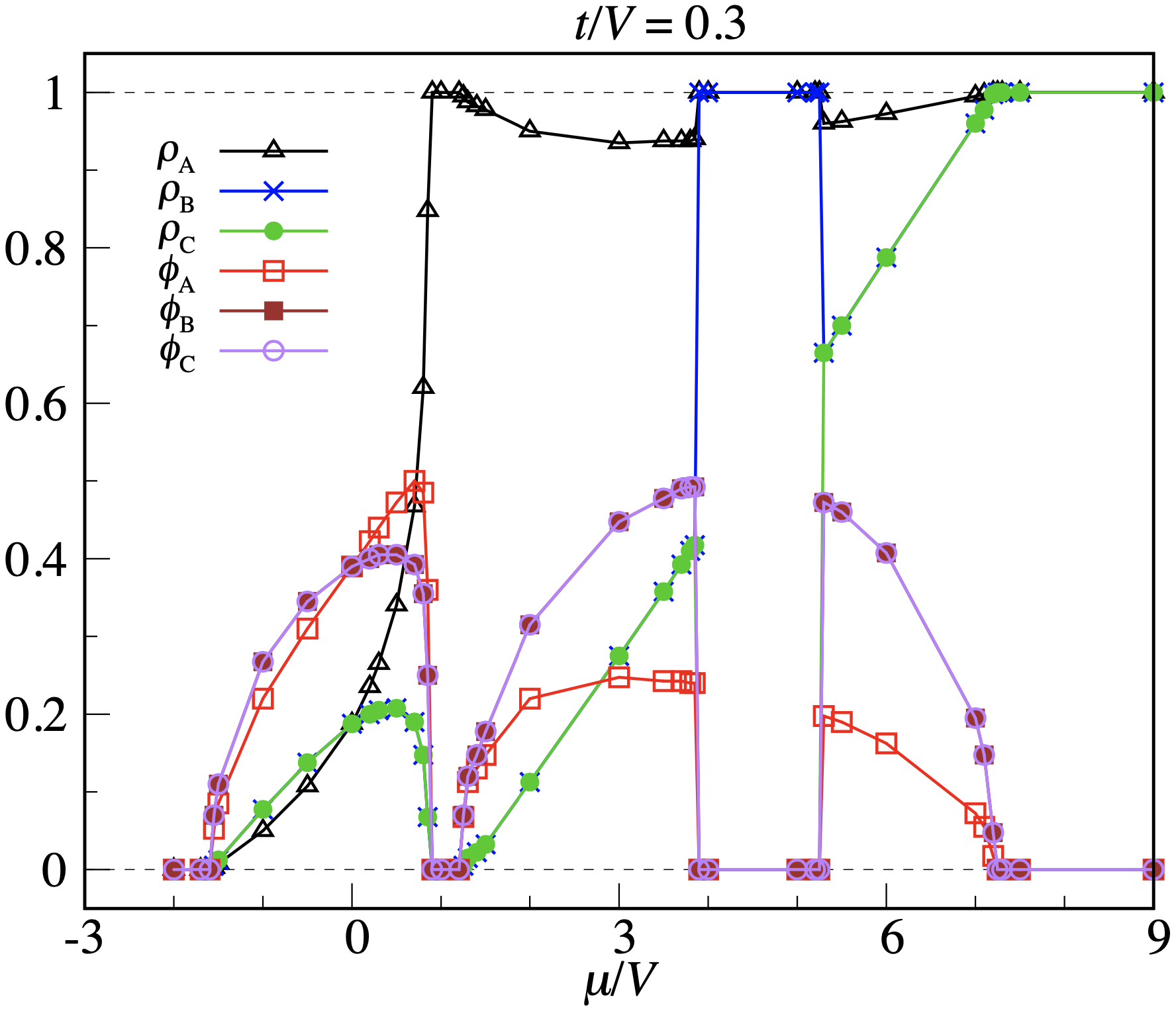}
\includegraphics[width=8.1cm]{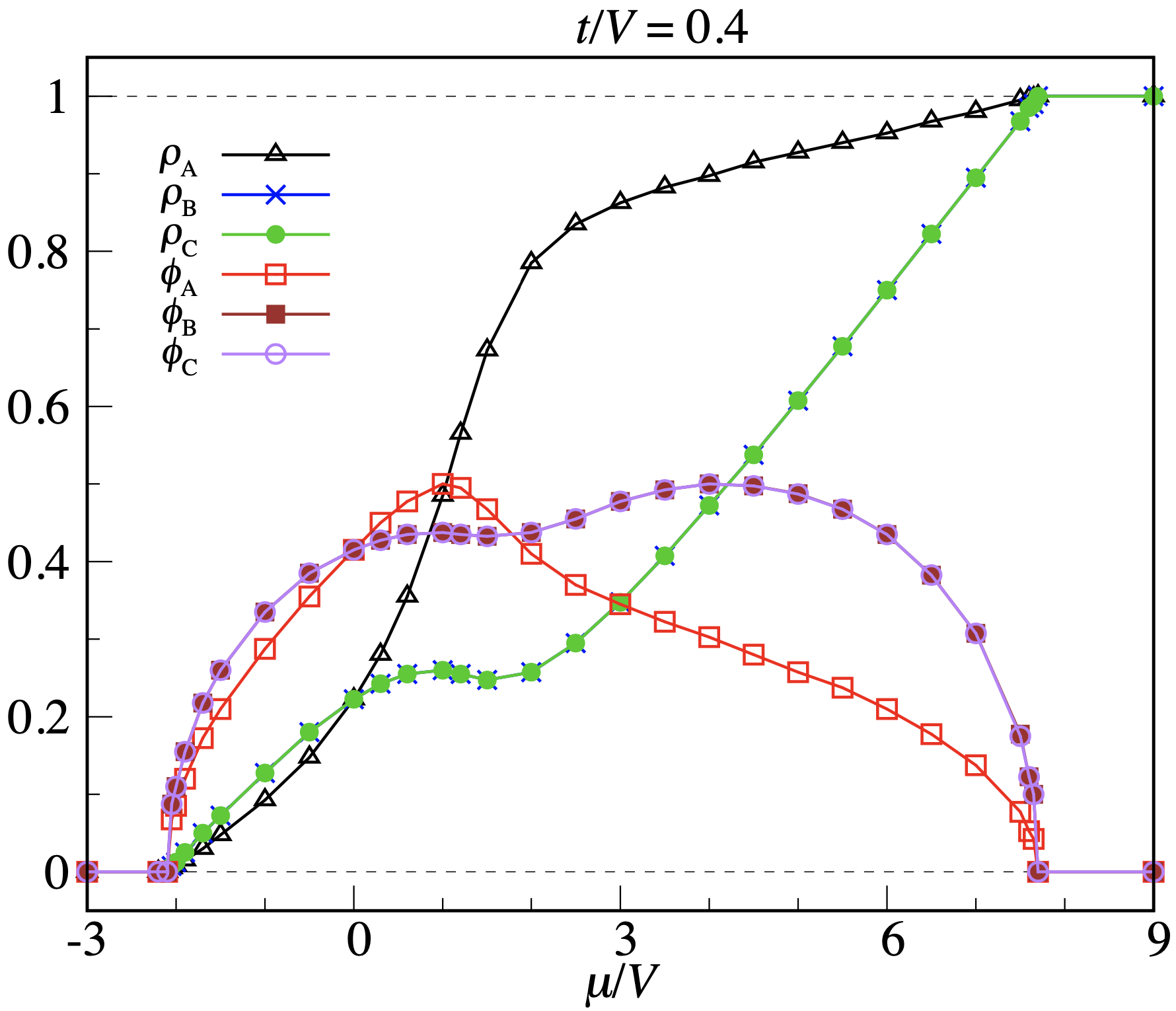}
\end{center}
\caption{Extended BH model on the TH graph. The DA order parameters are plotted as a function of $\mu$ for fixed $t$ (from top left to bottom right, $t/V=0.1,0.2,0.3,0.4$).}
\label{fig3}
\end{figure}

In Fig.\,3 we plot the order parameters as a function of $\mu$ for a number of $t$ values. The main message conveyed by the data is that, with the important exclusion of the OCT+TET phase, the number and superfluid density are the same on B and C. Moreover, some phase boundaries are continuous and other are first-order. The only exception is the boundary of the OCT phase, whose nature is twofold: while its descending branch is continuous, the ascending branch is first-order. The other continuous transitions are from ``empty'' to SS and from ``full'' to SS. Below, we perform a theoretical analysis of the functional dependence of $\mu$ on $t$ along each continuous-transition line, which is exact within the DA. Assuming full symmetry between B and C, we seek for solutions to Eqs.~(\ref{3-5}) and (\ref{3-6}) that match continuously with the values of the order parameters in the nearby insulating phase.

Near the transition line between ``empty'' and SS, every order parameter is close to zero. Expanding Eqs.~(\ref{3-5}) and (\ref{3-6}) near zero values we obtain:
\small
\be
\rho_{\rm A}\simeq\frac{16t^2\phi_{\rm B}^2}{(4V\rho_{\rm B}-\mu)^2}\,,\,\,\,
\rho_{\rm B}\simeq\frac{9t^2(\phi_{\rm A}+\phi_{\rm B})^2}{(3V\rho_{\rm A}+3V\rho_{\rm B}-\mu)^2}\,,\,\,\,\phi_{\rm A}\simeq\frac{4t\phi_{\rm B}}{4V\rho_{\rm B}-\mu}\,,\,\,\,
\phi_{\rm B}\simeq\frac{3t(\phi_{\rm A}+\phi_{\rm B})}{3V\rho_{\rm A}+3V\rho_{\rm B}-\mu}\,,
\label{3-8}
\ee
\normalsize
indicating that
\be
\rho_{\rm A}\simeq\phi_{\rm A}^2\,\,\,\,\,\,{\rm and}\,\,\,\,\,\,\rho_{\rm B}\simeq\phi_{\rm B}^2\,.
\label{3-9}
\ee
Plugging the latter equations in the last two Eqs.~(\ref{3-8}) and neglecting subdominant terms we arrive at two coupled equations for $\phi_{\rm A}$ and $\phi_{\rm B}$:
\be
\phi_{\rm A}+\frac{4t}{\mu}\phi_{\rm B}=0\,\,\,\,\,\,{\rm and}\,\,\,\,\,\,\frac{3t}{\mu}\phi_{\rm A}+\left(1+\frac{3t}{\mu}\right)\phi_{\rm B}=0\,.
\label{3-10}
\ee
In order that the linear set (\ref{3-10}) has non-zero solutions, the matrix of coefficients must have zero determinant:
\be
\mu^2+3t\mu-12t^2=0\,\,\,\,\,\,\longrightarrow\,\,\,\,\,\,\mu=-\frac{3+\sqrt{57}}{2}t\,.
\label{3-11}
\ee
The above equation gives the boundary line between ``empty'' and SS.

We may similarly expand Eqs.~(\ref{3-5}) and (\ref{3-6}) near $\rho_{\rm A}=\rho_{\rm B}=1$ and $\phi_{\rm A}=\phi_{\rm B}=0$, which are the order parameters in the ``full'' phase. We obtain:
\be
\rho_{\rm A}\simeq 1-\phi_{\rm A}^2\,\,\,\,\,\,{\rm and}\,\,\,\,\,\,\rho_{\rm B}\simeq 1-\phi_{\rm B}^2\,.
\label{3-12}
\ee
Inserting the above equations into the approximate expressions of $\phi_{\rm A}$ and $\phi_{\rm B}$ we arrive at two new coupled equations:
\be
\phi_{\rm A}+\frac{4t}{4V-\mu}\phi_{\rm B}=0\,\,\,\,\,\,{\rm and}\,\,\,\,\,\,\frac{3t}{6V-\mu}\phi_{\rm A}+\left(1+\frac{3t}{6V-\mu}\right)\phi_{\rm B}=0\,.
\label{3-13}
\ee
To have non-trivial solutions we need that
\small
\be
\mu^2-(10V+3t)\mu+24V^2+12Vt-12t^2=0\,\,\,\,\,\,\longrightarrow\,\,\,\,\,\,\mu=\frac{10V+3t+\sqrt{4V^2+12Vt+57t^2}}{2}\,,
\label{3-14}
\ee
\normalsize
giving the boundary between ``full'' and SS.

Finally, near the descending branch of the OCT boundary we have solutions to Eqs.~(\ref{3-5}) and (\ref{3-6}) that are close to $\rho_{\rm A}=1,\rho_{\rm B}=\phi_{\rm A}=\phi_{\rm B}=0$. We easily find:
\be
\rho_{\rm A}\simeq 1-\phi_{\rm A}^2\,\,\,\,\,\,{\rm and}\,\,\,\,\,\,\rho_{\rm B}\simeq\phi_{\rm B}^2\,.
\label{3-15}
\ee
Inserting the latter equations into the expressions of $\phi_{\rm A}$ and $\phi_{\rm B}$ we obtain a new set of linear equations:
\be
\phi_{\rm A}-\frac{4t}{\mu}\phi_{\rm B}=0\,\,\,\,\,\,{\rm and}\,\,\,\,\,\,\frac{3t}{3V-\mu}\phi_{\rm A}-\left(1-\frac{3t}{3V-\mu}\right)\phi_{\rm B}=0\,.
\label{3-16}
\ee
We have non-trivial solutions provided that
\be
\mu^2-(3V-3t)\mu+12t^2=0\,\,\,\,\,\,\longrightarrow\,\,\,\,\,\,\mu_\pm=\frac{3V-3t\pm\sqrt{9V^2-18Vt-39t^2}}{2}\,.
\label{3-17}
\ee
While $\mu_+$ describes the descending branch of the OCT-SS boundary, the solution $\mu_-$ is discarded since it corresponds to a (virtual) continuous transition from OCT to SS that is preempted by a first-order transition occurring close to $\mu_-$. Observe that the square root in (\ref{3-17}) only exists for $t\le(4\sqrt{3}-3)V/13=0.3021\ldots V$, which then represents the abscissa $t_c$ of the (tri)critical point (the ordinate being $\mu_c=(3V-3t_c)/2=1.0467\ldots V$). 

\subsection{TH model: exact zero-temperature analysis}

For the TH model, the dimensionality of the Hilbert space ($2^{14}=16384$) is small enough that we can compute a few exact energy eigenvalues and relative eigenstates in affordable time. To this aim we represent the Hamiltonian on the Fock basis $\{\ve{x_1,x_2,\ldots,x_{14}}\}$ (with $x_i=0$ or 1) and diagonalize the ensuing matrix numerically. In particular, the ground state $\ve g$ and its eigenvalue, the grand potential $\Omega$, can be mapped as a function of $t$ and $\mu$.

Once $\ve g$ has been determined, we calculate the average occupancies of A, B, and C nodes (corresponding to the MF parameters $\rho_{\rm A},\rho_{\rm B}$, and $\rho_{\rm C}$), the average value of $a_i$, and the {\em superfluid density} $\rho_{\rm SF}$ (see, e.g., Refs.~\cite{vanOosten,Yamamoto2}). The latter quantity reads:
\be
\rho_{\rm SF}\equiv\frac{1}{14}\meb g{\widetilde{a}_{\bf 0}^\dagger\widetilde{a}_{\bf 0}\big|g}=\frac{1}{14^2}\sum_{i,j=1}^{14}\meb g{a_i^\dagger a_j\big|g}\,,
\label{3-18}
\ee
where $\widetilde{a}_{\bf 0}=(1/\sqrt{14})\sum_{i=1}^{14}a_i$ is the zero-momentum field operator. Observe that, in a large lattice of $M$ sites, $\big<\widetilde{a}_{\bf 0}^\dagger\widetilde{a}_{\bf 0}\big>=N_0$ is the average number of condensate particles, hence $\rho_{\rm SF}=N_0/M$ is the condensate density.

%
%
\begin{figure}
\begin{center}
\includegraphics[width=15cm]{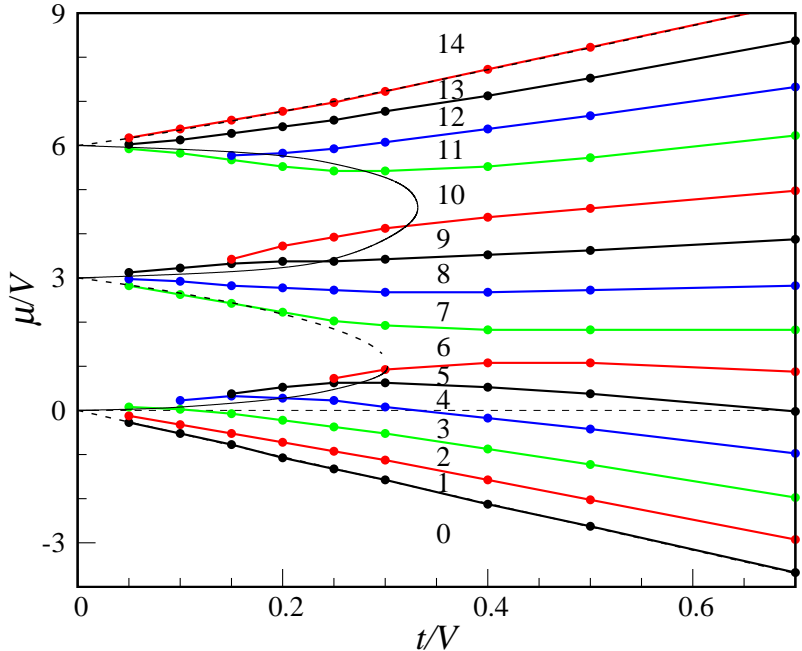}
\end{center}
\caption{TH model, exact diagonalization vs. MF results. In this ``phase diagram'', any difference between distinct phases is blurred. The thin dashed and continuous lines are the MF transition lines. The thick lines through the dots separate sectors of the phase diagram where $N$ and other averages are constant.}
\label{fig4}
\end{figure}

%
%
\begin{figure}
\begin{center}
\includegraphics[width=8.1cm]{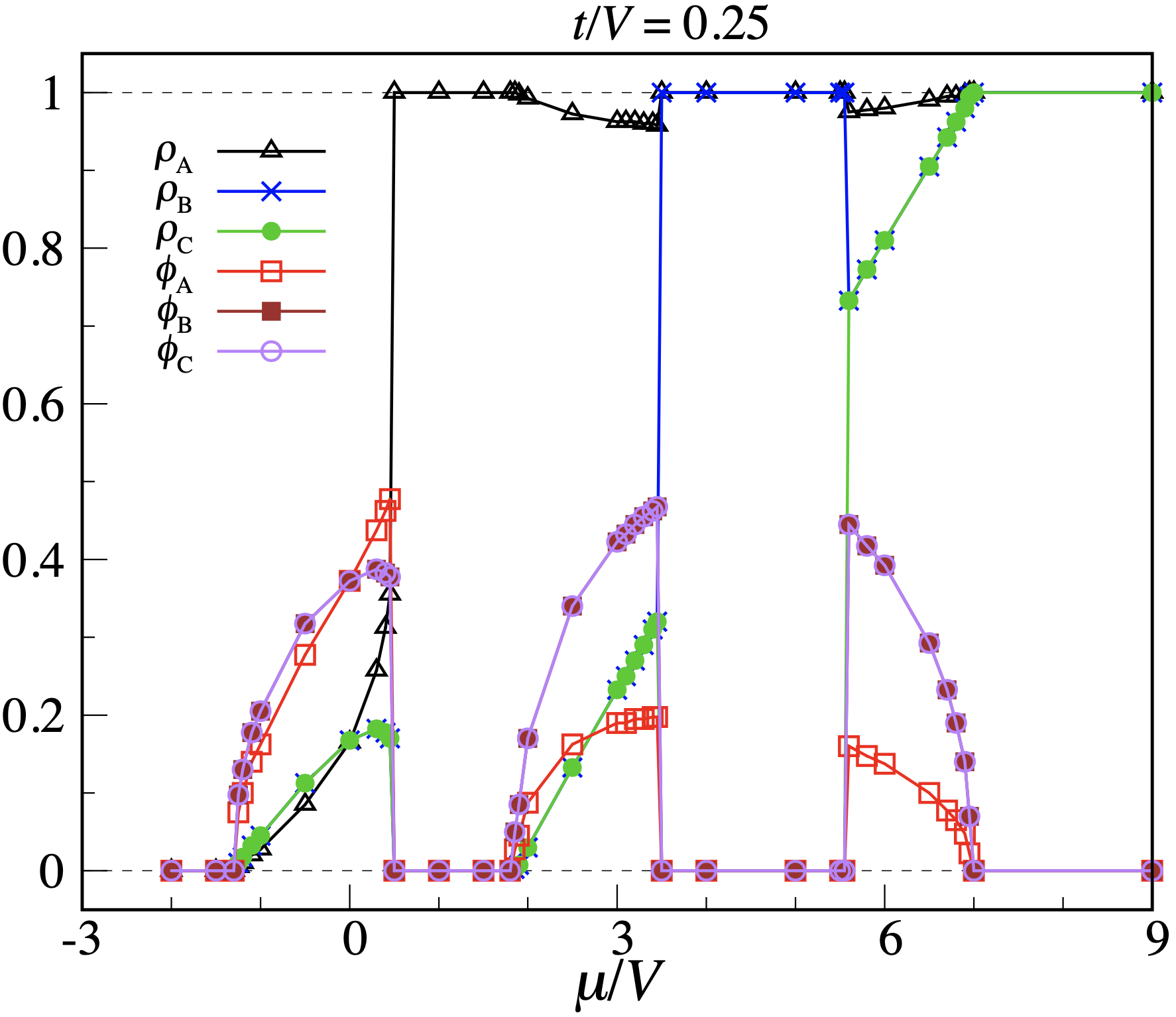}
\includegraphics[width=8.1cm]{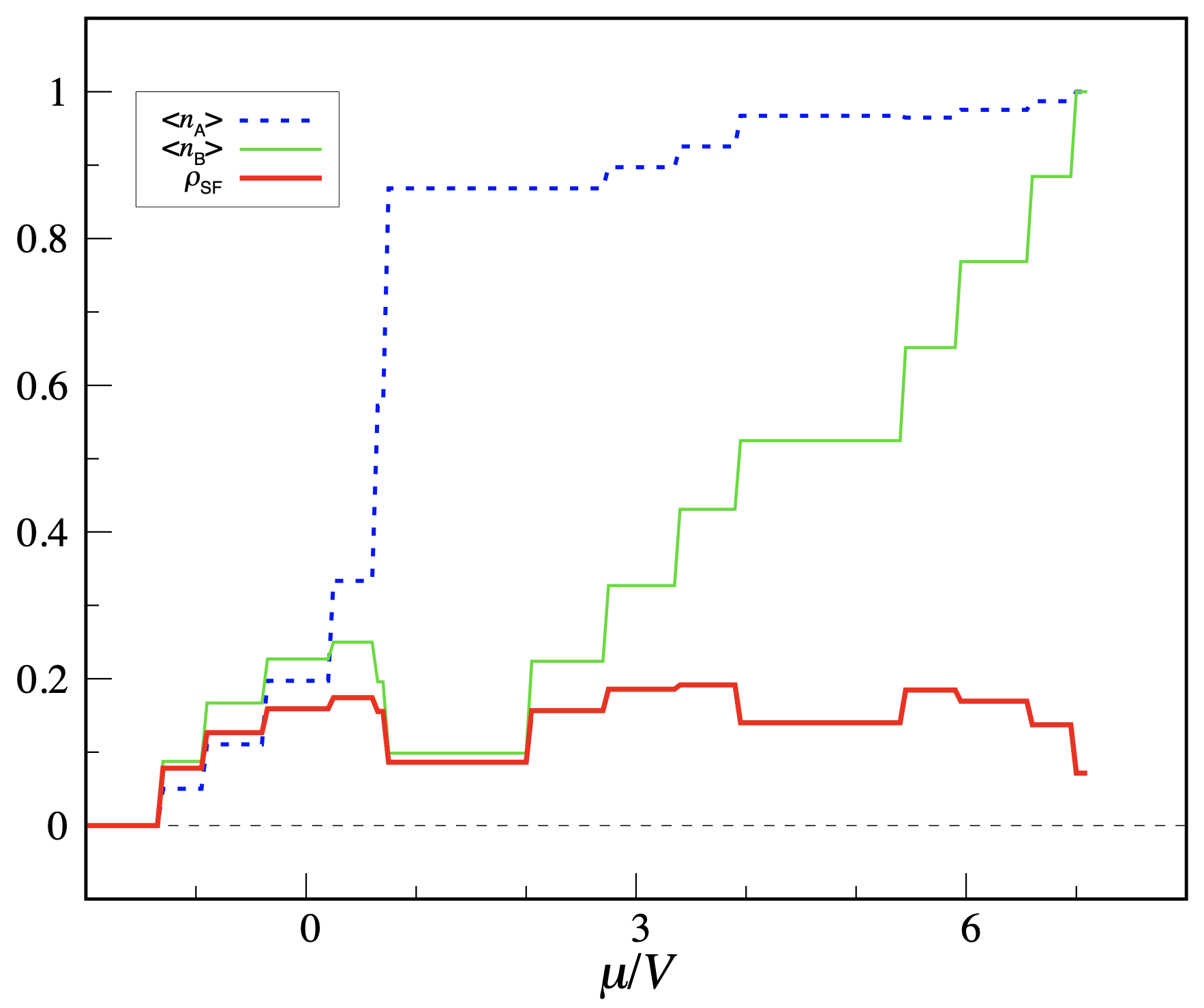}
\end{center}
\caption{Extended BH model on the TH graph: order parameters plotted as a function of $\mu$ for $t=0.25\,V$. Left: DA results (with the only exception of the OCT+TET ``phase'', $\rho_{\rm B}$ and $\rho_{\rm C}$ are practically equal). Right: exact results ($\langle n_{\rm A}\rangle$, dotted blue line; $\langle n_{\rm B}\rangle$, thin green line; $\rho_{\rm SF}$, red thick line). Here $\langle n_{\rm B}\rangle=\langle n_{\rm C}\rangle$ is an outcome of diagonalization.}
\label{fig5}
\end{figure}

In doing the computations, we find a perfect symmetry between the vertex subsets B and C, also in the putative OCT+TET region. The only exception is $t=0$, where the B-C symmetry is broken and the node occupancies are the same as in MF theory. Since the Hamiltonian commutes with the total number of particles $\sum_ia_i^\dagger a_i$, the $(t,\mu)$ plane is divided in sectors where the number of particles takes a constant integer value $N$, from 0 to 14. As expected, $N=0$ in the ``empty'' phase and $N=14$ in the ``full'' phase. In the $N$-sector, the only non-zero Fourier coefficients of $\ve g$ are those relative to basis states with $\sum_ix_i=N$. The resulting ``phase diagram'' is plotted in Fig.\,4. In stark contrast with the MF phase diagram (Fig.\,2), there are no sharp phase boundaries. This is more clearly visible in Fig.\,5, where we make a comparison in terms of order parameters between exact diagonalization and MF theory for $t=0.25\,V$. The exact $\mu$ evolution of $\langle n_{\rm A}\rangle$ and $\langle n_{\rm B}\rangle=\langle n_{\rm C}\rangle$ roughly traces the MF curves, except for the $N=10$ sector --- corresponding to the crossing of the OCT+TET region --- where instead $\langle n_{\rm B}\rangle\ne\langle n_{\rm C}\rangle$.

Another difference with MF theory is in the ground-state average of $a_i$, which is identically zero. In fact, we have already commented in Ref.~\cite{Prestipino6} that the right quantity to look at is the superfluid density $\rho_{\rm SF}$ (red curve in Fig.\,5b), which indeed compares well with $\phi_i^2$. In particular, $\rho_{\rm SF}$ drops to a minimum where $\phi_i$ vanishes, i.e., in the $\mu$ ranges pertaining to the insulating phases. The non-zero value of $\rho_{\rm SF}$ in these phases is a finite-size effect. A slightly larger value of $\rho_{\rm SF}$ in the OCT+TET region could be the result of a free circulation of particles within the cubic sites.

%
%
\begin{figure}
\begin{center}
\includegraphics[width=15cm]{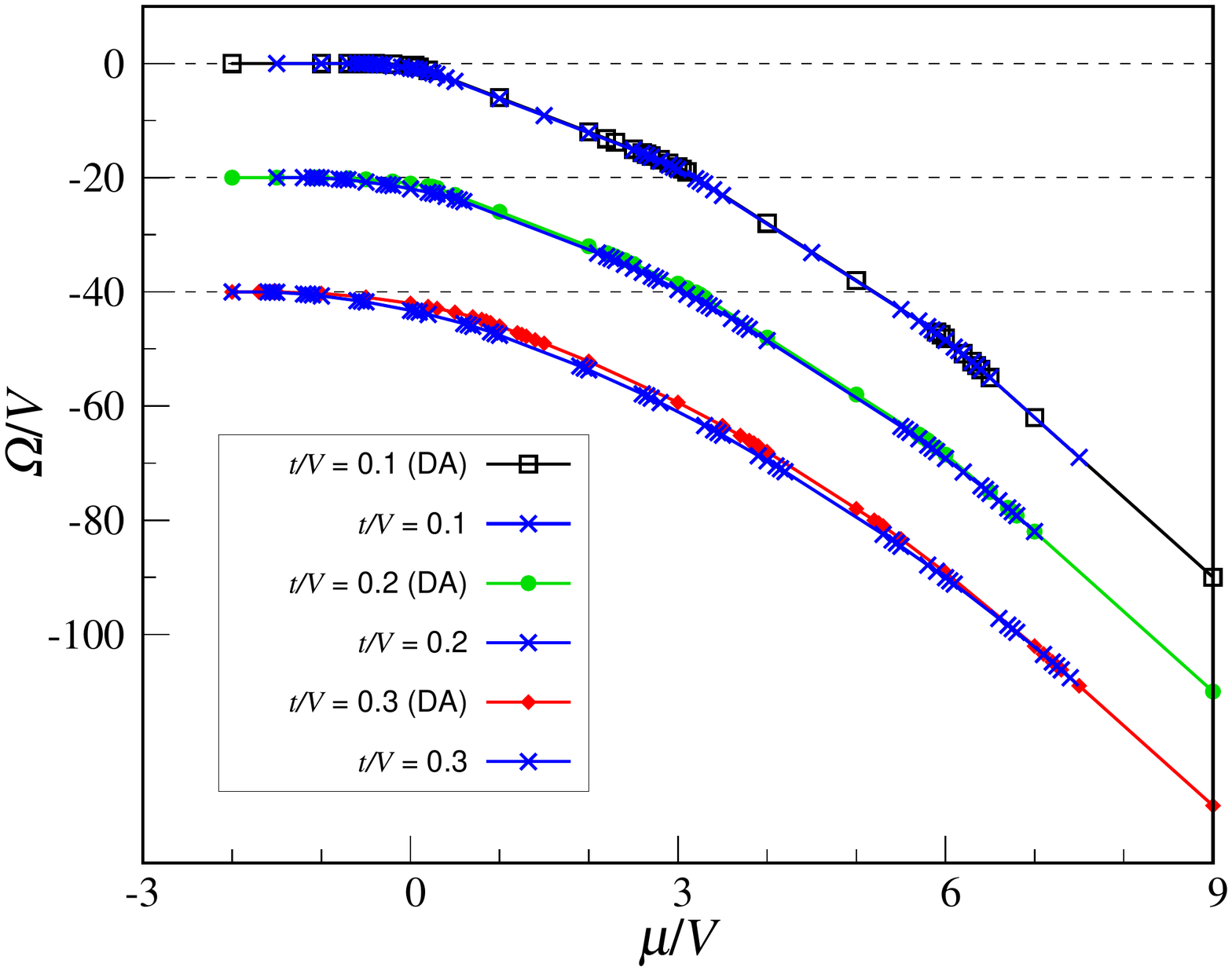}
\end{center}
\caption{Extended BH model on the TH graph. The DA grand potential (black squares, green dots, and red diamonds) is plotted as a function of $\mu$ for fixed $t$ (for $t/V=0.1,0.2,0.3$), and compared with the exact value (blue crosses) obtained from Hamiltonian diagonalization. To help visualization, data for $t/V=0.2$ (0.3) have been shifted downwards by 20 (40).}
\label{fig6}
\end{figure}

%
%
\begin{figure}
\begin{center}
\includegraphics[width=15cm]{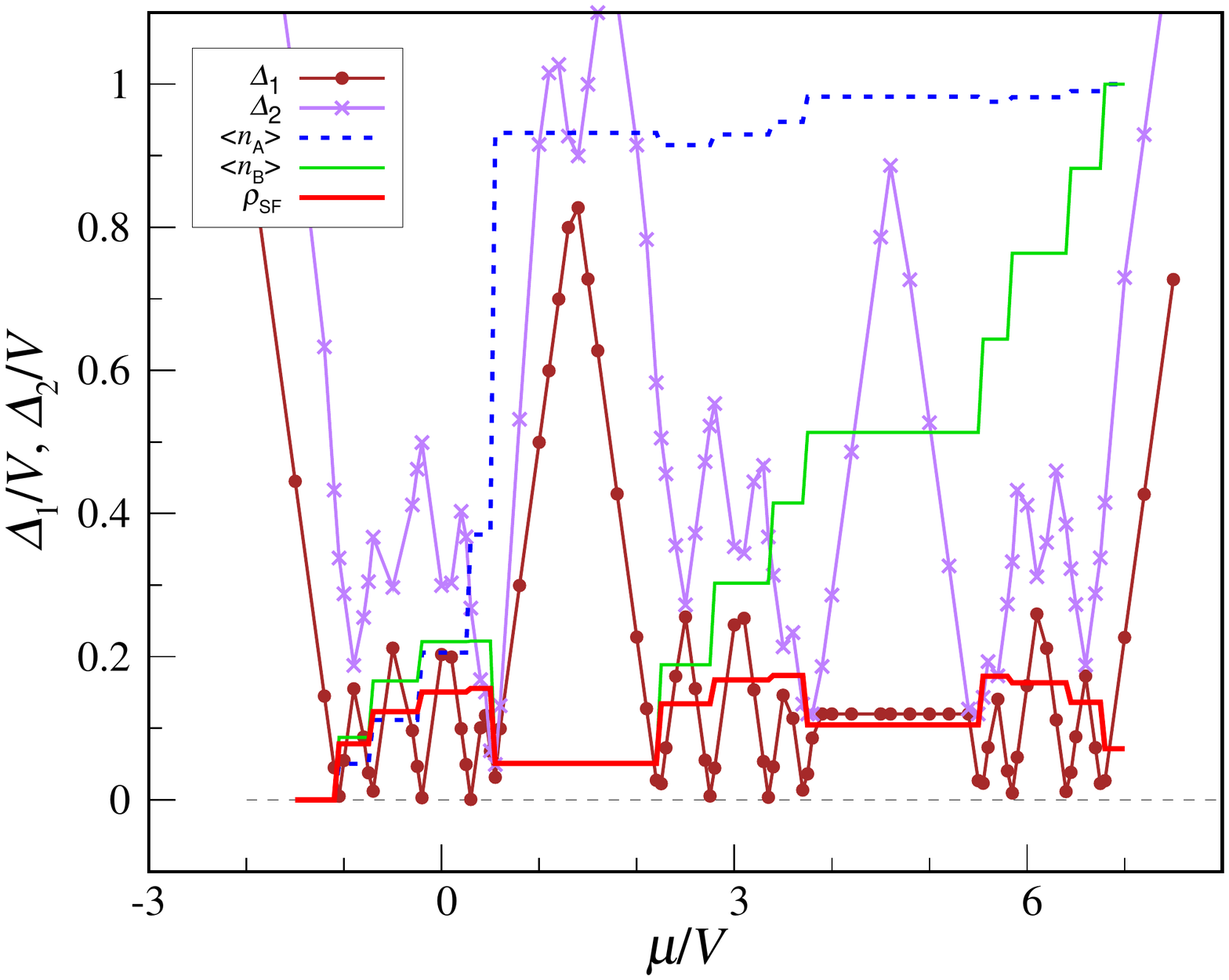}
\end{center}
\caption{Extended BH model on the TH graph. The first and second gaps are plotted as a function of $\mu$ for $t=0.20\,V$. For a better reading of the data, also the average occupancies $\langle n_{\rm A}\rangle,\langle n_{\rm B}\rangle$ and the superfluid density $\rho_{\rm SF}$ have been reported ($\langle n_{\rm A}\rangle$, dotted blue line; $\langle n_{\rm B}\rangle$, thin green line; $\rho_{\rm SF}$, red thick line).}
\label{fig7}
\end{figure}

To get a flavor of the quality of MF theory, we may look at Fig.\,6 where the exact and approximate grand potentials are plotted as a function of $\mu$ for a few $t$ values. We see that MF data lie systematically above the exact values, as should be expected for a variational estimate based on the Gibbs-Bogoliubov inequality (see Appendix B). We also generally confirm that
\be
\frac{\partial\Omega}{\partial\mu}=-N
\label{3-19}
\ee
and that MF theory worsens with increasing $t$, as already evident in Fig.\,4.

A distinguishing feature of an insulating phase is a non-zero energy gap, in contrast to the zero gap of a superfluid/supersolid phase (see, e.g., \cite{Bloch}). To see whether this is confirmed in our system, in addition to the lowest energy eigenvalue $\Omega$, we have also computed the second ($\Omega_2$) and the third energy eigenvalue ($\Omega_3$), which define the first and second gaps, $\Delta_1=\Omega_2-\Omega$ and $\Delta_2=\Omega_3-\Omega$. These two quantities are plotted in Fig.\,7 as a function of $\mu$ for $t=0.20\,V$. We see that both gaps are larger in the insulating phases than in the SS regions; as a rule, $\Delta_1$ is wider the larger is the distance in chemical potential from the line separating two consecutive sectors in Fig.\,4. The non-monotonic behavior of $\Delta_1$ with $\mu$ has a simple explanation: while the less-costly excitation is hole-like on the low-$\mu$ side of a sector, it is particle-like on the high-$\mu$ side. Looking more closely to the data, we indeed realize that
\be
\frac{\partial\Delta_1}{\partial\mu}=\pm 1\,,
\label{3-20}
\ee
meaning that the first excited state, which is generally non-degenerate, is a linear combination of basis states with one particle more or less than those composing the ground state. Only for $N=10$ the above derivative is zero, meaning that the first excited state is, like the ground state, a linear combination of basis states having $\sum_ix_i=10$.

\subsection{TH model: MF theory in the spin representation}

It is instructive to see how the same DA results at $T=0$ can be recovered by working in the representation where the extended BH model with $U=+\infty$ is mapped onto a spin-1/2 Hamiltonian. We recap in Appendix D the exact terms of this correspondence, which goes back to a paper by Matsubara and Matsuda~\cite{Matsubara}. Below, we treat the case of the TH model.

The TH vertices are of three types: six octahedral nodes (A), four tetrahedral-1 nodes (B), and 4 tetrahedral-2 nodes (C). Depending on the sites involved, the number of distinct nearest-neighbor pairs is either 0 (AA-, BB-, and CC-type) or 12 (AB-, AC-, and BC-type). Starting from the BH Hamiltonian in the spin representation,
\be
H=\sum_{\left<i,j\right>}\left[VS_i^zS_j^z-2t\big(S_i^xS_j^x+S_i^yS_j^y\big)\right]+\frac{V}{2}\sum_{\left<i,j\right>}\left(S_i^z+S_j^z+\frac{1}{2}\right)-\mu\sum_i\left(S_i^z+\frac{1}{2}\right)\,,
\label{3-21}
\ee
the MF energy is obtained by replacing the quantum spins with classical spins of magnitude $1/2$, further assuming the same spin vector in all nodes of same type:
\ba
E_S&=&3V\big[\cos\theta_{\rm A}\cos\theta_{\rm B}+\cos\theta_{\rm A}\cos\theta_{\rm C}+\cos\theta_{\rm B}\cos\theta_{\rm C}
\nonumber \\
&-&\Delta(\sin\theta_{\rm A}\sin\theta_{\rm B}+\sin\theta_{\rm A}\sin\theta_{\rm C}+\sin\theta_{\rm B}\sin\theta_{\rm C})\big]
\nonumber \\
&+&(6V-3\mu)\cos\theta_{\rm A}+(6V-2\mu)\cos\theta_{\rm B}+(6V-2\mu)\cos\theta_{\rm C}+9V-7\mu\,,
\label{3-22}
\ea
where $\Delta=2t/V$ and, for example, $\theta_{\rm A}$ is the angle between ${\bf S}_{\rm A}$ and $\hat{\bf z}$. With no loss of generality, we can assume that in the minimum-energy configurations the spins are all lying in the $x$-$z$ plane.

We first examine the minimum-energy states for $t=0$, where every spin points in the $z$ direction:
\ba
&&\downarrow\,\downarrow\,\downarrow\,\,:\,\,\,\,\,\,E_S=0\,\,\,({\rm ``empty"})\,;
\nonumber \\
&&\uparrow\,\uparrow\,\uparrow\,\,:\,\,\,\,\,\,E_S=36V-14\mu\,\,\,({\rm ``full"})\,;
\nonumber \\
&&\uparrow\,\downarrow\,\downarrow\,\,:\,\,\,\,\,\,E_S=-6\mu\,\,\,({\rm OCT})\,;
\nonumber \\
&&\uparrow\,\uparrow\,\downarrow\,\,\,{\rm and}\,\,\,\uparrow\,\downarrow\,\uparrow\,\,:\,\,\,\,\,\,E_S=12V-10\mu\,\,\,({\rm OCT+TET})\,;
\nonumber \\
&&\downarrow\,\uparrow\,\uparrow\,\,:\,\,\,\,\,\,E_S=12V-8\mu\,\,\,({\rm CUB})\,.
\label{3-23}
\ea
The above spin energies are equal to the grand-potential values as previously determined for the TH model, hence the same sequence of $t=0$ phases occurs as a function of $\mu$.

For a supersolid phase with $\theta_{\rm A}\ne\theta_{\rm B}=\theta_{\rm C}$, the total energy takes the form
\ba
E_S({\rm SS})&=&3V\big[2\cos\theta_{\rm A}\cos\theta_{\rm B}+\cos^2\theta_{\rm B}-\Delta(2\sin\theta_{\rm A}\sin\theta_{\rm B}+\sin^2\theta_{\rm B})\big]
\nonumber \\
&+&(6V-3\mu)\cos\theta_{\rm A}+2(6V-2\mu)\cos\theta_{\rm B}+9V-7\mu\,.
\label{3-24}
\ea
Assume that the system is initially in the OCT phase ($\cos\theta_{\rm A}=1,\cos\theta_{\rm B}=-1$). A continuous transition to SS occurs as the point of absolute minimum energy moves to $\cos\theta_{\rm A}\lesssim 1,\cos\theta_{\rm B}\gtrsim -1$. Expanding $\Delta E_S=E_S({\rm SS})-E_S({\rm OCT})$ around $\theta_{\rm A}=0$ and $\theta_{\rm B}=\pi$ we obtain:
\be
\Delta E_S\simeq 3V\big[\theta_{\rm A}^2+\Delta(2\theta_{\rm A}\theta_{\rm B}'-\theta_{\rm B}^{\prime 2})\big]-\frac{1}{2}(6V-3\mu)\theta_{\rm A}^2+(6V-2\mu)\theta_{\rm B}^{\prime 2}
\label{3-25}
\ee
with $\theta_{\rm B}'=\theta_{\rm B}-\pi$. A non-zero stationary point occurs when the Hessian becomes negative. This requires
\be
2\mu^2-3(2V-\Delta V)\mu+6\Delta^2V^2>0\,,
\label{3-26}
\ee
yielding $\mu\lesssim\mu_-$ or $\mu\gtrsim\mu_+$ with
\be
\mu_\pm=\frac{3V-3t\pm\sqrt{9V^2-18Vt-39t^2}}{2}\,.
\label{3-27}
\ee
The transition to SS for $\mu=\mu_-$ is actually preempted by a first-order transition. Notice that Eq.~(\ref{3-27}) is equivalent to Eq.~(\ref{3-17}).

A continuous transition from ``filled'' to SS occurs when the absolute minimum of $E_S$ moves from $\cos\theta_{\rm A}=1,\cos\theta_{\rm B}=1$ to $\cos\theta_{\rm A}\lesssim 1,\cos\theta_{\rm B}\lesssim 1$. The relative energy between ``filled'' and SS is
\be
\Delta E_S\equiv E_S({\rm SS})-E_S(``{\rm filled}")\simeq 3V\big[-\theta_{\rm A}^2-2\Delta\theta_{\rm A}\theta_{\rm B}-(2+\Delta)\theta_{\rm B}^2\big]-\frac{1}{2}(6V-3\mu)\theta_{\rm A}^2-(6V-2\mu)\theta_{\rm B}^2\,.
\label{3-28}
\ee
A non-zero stationary point only occurs for
\be
2\mu^2-(20V+3\Delta V)\mu+12V^2(4+\Delta)-6\Delta^2 V^2<0\,,
\label{3-29}
\ee
which is certainly satisfied for $\mu\lesssim\mu_+$ with
\be
\mu_+=\frac{10V+3t+\sqrt{4V^2+12Vt+57t^2}}{2}\,,
\label{3-30}
\ee
coincident with Eq.~(\ref{3-14}).

Finally, we observe a continuous transition from ``empty'' to SS when the absolute minimum of $E_S$ moves from $\cos\theta_{\rm A}=-1,\cos\theta_{\rm B}=-1$ to $\cos\theta_{\rm A}\gtrsim -1,\cos\theta_{\rm B}\gtrsim -1$. Upon defining $\theta_{\rm A}'=\theta_{\rm A}-\pi$ and $\theta_{\rm B}'=\theta_{\rm B}-\pi$, we obtain
\be
\Delta E_S\equiv E_S({\rm SS})-E_S(``{\rm empty}")\simeq -3V\big[\theta_{\rm A}^{\prime 2}+2\Delta\theta_{\rm A}'\theta_{\rm B}'+(2+\Delta)\theta_{\rm B}^{\prime 2}\big]+\frac{1}{2}(6V-3\mu)\theta_{\rm A}^{\prime 2}+(6V-2\mu)\theta_{\rm B}^{\prime 2}\,.
\label{3-31}
\ee
A non-zero stationary point only exists if the Hessian of (\ref{3-31}) is negative, that is for
\be
2\mu^2+3\Delta V\mu-6\Delta^2V^2<0\,,
\label{3-32}
\ee
which is certainly satisfied for $\mu\gtrsim\mu_-$ with
\be
\mu_-=-\frac{3+\sqrt{57}}{2}t\,.
\label{3-33}
\ee
The above equation is the same as Eq.~(\ref{3-11}).

For a general analysis of the characteristics of the B-C symmetric case we need to express the MF energy $E_S$ as a function of four order parameters. To this aim one observes that
\be
\frac{1}{2}\cos\theta_{{\rm A},{\rm B}}=\rho_{{\rm A},{\rm B}}-\frac{1}{2}\,\,\,\,\,\,{\rm and}\,\,\,\,\,\,\frac{1}{2}\sin\theta_{{\rm A},{\rm B}}=\phi_{{\rm A},{\rm B}}\,,
\label{3-34}
\ee
which can be combined to give
\be
\left(\rho_{{\rm A},{\rm B}}-\frac{1}{2}\right)^2+\phi_{{\rm A},{\rm B}}^2=\frac{1}{4}\,.
\label{3-35}
\ee
Eliminating $\phi_{{\rm A},{\rm B}}$ in favor of $\rho_{{\rm A},{\rm B}}$ through Eq.~(\ref{3-35}), the MF energy becomes
\ba
E_S&=&24V\rho_{\rm A}\rho_{\rm B}+12V\rho_{\rm B}^2-12t\sqrt{1-4(\rho_{\rm A}-1/2)^2}\sqrt{1-4(\rho_{\rm B}-1/2)^2}
\nonumber \\
&-&6t+24t\left(\rho_{\rm B}-\frac{1}{2}\right)^2-6\mu\rho_{\rm A}-8\mu\rho_{\rm B}\,,
\label{3-36}
\ea
whose stationary points obey the following equations:
\be
\frac{\partial E_S}{\partial\rho_{\rm A}}=0\,\,\,\,\,\,{\rm and}\,\,\,\,\,\,\frac{\partial E_S}{\partial\rho_{\rm B}}=0\,.
\label{3-37}
\ee
The former equation leads to
\be
\rho_{\rm A}-\frac{1}{2}=-\frac{4V\rho_{\rm B}-\mu}{\left(\frac{8t\phi_{\rm B}}{\phi_{\rm A}}\right)}\,.
\label{3-38}
\ee
Plugging this equation in (\ref{3-35}) we arrive at
\be
\phi_{\rm A}=\frac{4t\phi_{\rm B}}{\sqrt{(4V\rho_{\rm B}-\mu)^2+64t^2\phi_{\rm B}^2}}\,.
\label{3-39}
\ee
Inserting the latter equation back in (\ref{3-38}) we obtain
\be
\rho_{\rm A}=\frac{1}{2}-\frac{4V\rho_{\rm B}-\mu}{2\sqrt{(4V\rho_{\rm B}-\mu)^2+64t^2\phi_{\rm B}^2}}\,.
\label{3-40}
\ee
By a similar line of thought, from the second of Eqs.\,(\ref{3-37}) we arrive at
\be
\phi_{\rm B}=\frac{3t(\phi_{\rm A}+\phi_{\rm B})}{\sqrt{(3V\rho_{\rm A}+3V\rho_{\rm B}-\mu)^2+36t^2(\phi_{\rm A}+\phi_{\rm B})^2}}
\label{3-41}
\ee
and
\be
\rho_{\rm B}=\frac{1}{2}-\frac{3V\rho_{\rm A}+3V\rho_{\rm B}-\mu}{2\sqrt{(3V\rho_{\rm A}+3V\rho_{\rm B}-\mu)^2+36t^2(\phi_{\rm A}+\phi_{\rm B})^2}}\,.
\label{3-42}
\ee
Equations (\ref{3-39})-(\ref{3-42}) exactly coincide with Eqs.~(\ref{3-5}) and (\ref{3-6}) when perfect symmetry is assumed between B and C.

\subsection{PD model}

We conclude with the DA analysis at $T=0$ of a system of hard-core bosons on the PD graph, following the same lines of reasoning as in Section 3.A. Looking at Fig.\,1b, the 32 nodes of the PD graph are naturally classified as icosahedral (12) or dodecahedral (20). In fact, the existing repulsion between NN particles recommends to distinguish between dodecahedral nodes of cubic (8) and co-cubic type (12)~\cite{Prestipino6}. Hence, we have three types of PD vertices: icosahedral (A), cubic (B), and co-cubic (C). Upon considering that
\ba
&&F_{\rm A}=2\phi_{\rm B}+3\phi_{\rm C}\,,\,\,\,F_{\rm B}=3\phi_{\rm A}+3\phi_{\rm C}\,,\,\,\,F_{\rm C}=3\phi_{\rm A}+2\phi_{\rm B}+\phi_{\rm C}\,;
\nonumber \\
&&R_{\rm A}=2\rho_{\rm B}+3\rho_{\rm C}\,,\,\,\,R_{\rm B}=3\rho_{\rm A}+3\rho_{\rm C}\,,\,\,\,R_{\rm C}=3\rho_{\rm A}+2\rho_{\rm B}+\rho_{\rm C}\,,
\label{3-43}
\ea
the MF Hamiltonian reads:
\small
\ba
H_{\rm DA}&=&E_0-12t\left[(2\phi_{\rm B}+3\phi_{\rm C})a_{\rm A}^\dagger+(2\phi_{\rm B}^*+3\phi_{\rm C}^*)a_{\rm A}\right]-8t\left[(3\phi_{\rm A}+3\phi_{\rm C})a_{\rm B}^\dagger+(3\phi_{\rm A}^*+3\phi_{\rm C}^*)a_{\rm B}\right]
\nonumber \\
&-&12t\left[(3\phi_{\rm A}+2\phi_{\rm B}+\phi_{\rm C})a_{\rm C}^\dagger+(3\phi_{\rm A}^*+2\phi_{\rm B}^*+\phi_{\rm C}^*)a_{\rm C}\right]+12(2V\rho_{\rm B}+3V\rho_{\rm C}-\mu)n_{\rm A}
\nonumber \\
&+&8(3V\rho_{\rm A}+3V\rho_{\rm C}-\mu)n_{\rm B}+12(3V\rho_{\rm A}+2V\rho_{\rm B}+V\rho_{\rm C}-\mu)n_{\rm C}
\label{3-44}
\ea
\normalsize
with
\ba
E_0&=&12t\left[2(\phi_{\rm A}^*\phi_{\rm B}+\phi_{\rm A}\phi_{\rm B}^*)+3(\phi_{\rm A}^*\phi_{\rm C}+\phi_{\rm A}\phi_{\rm C}^*)+2(\phi_{\rm B}^*\phi_{\rm C}+\phi_{\rm B}\phi_{\rm C}^*)\right]+12t|\phi_{\rm C}|^2
\nonumber \\
&-&12V\left(2\rho_{\rm A}\rho_{\rm B}+3\rho_{\rm A}\rho_{\rm C}+2\rho_{\rm B}\rho_{\rm C}\right)-6V\rho_{\rm C}^2\,.
\label{3-45}
\ea
Like for the TH model, the stable insulating phases at $t=0$ can be identified by looking at the elements of the diagonal matrix representing (\ref{3-44}) on the canonical basis $\ve{x_{\rm A},x_{\rm B},x_{\rm C}}$. A calculation similar to the one in Section 3.A leads to the following table:
\ba
\begin{tabular}{cccc}
$\mu\,\,{\rm range}$\qquad\qquad & {\rm grand\,\,potential}\qquad\qquad & {\rm ground\,\,state}\qquad\qquad & phase\\
\hline
$\mu\le 0$\,:\qquad\qquad & $0$\qquad\qquad & $\ve{0,0,0}$\qquad\qquad & {\rm ``empty''}\\
$0\le\mu\le 3V$\,:\qquad\qquad & $-12\mu$\qquad\qquad & $\ve{1,0,0}$\qquad\qquad & {\rm ICO}\\
$3V\le\mu\le(9/2)V$\,:\qquad\qquad & $24V-20\mu$\qquad\qquad & $\ve{1,1,0}$\qquad\qquad & {\rm ICO+CUB}\\
$(9/2)V\le\mu\le 6V$\,:\qquad\qquad & $42V-24\mu$\qquad\qquad & $\ve{1,0,1}$\qquad\qquad & {\rm ICO+CCO}\\
$\mu\ge 6V$\,:\qquad\qquad & $90V-32\mu$\qquad\qquad & $\ve{1,1,1}$\qquad\qquad & {\rm ``full''}
\nonumber\\
\end{tabular}
\ea
In the above list of phases, ``ICO'' is the phase where all the icosahedral nodes are occupied ($N=12$ particles in total); ``ICO+CUB'' is the phase where A and B are filled ($N=20$); ``ICO+CCO'' is the phase where A and C are filled ($N=24$); finally, ``full'' is the phase where there is one particle at each node ($N=32$). Notice that a hypothetical ICO+TET phase ($\Omega=12V-16\mu$) would only be stable at the single point $\mu=3V$ and here degenerate with ICO and ICO+CUB. Should we have opted for a notion of nearness based on spatial proximity, we would have obtained a stable DOD phase (i.e., one with all the dodecahedral nodes occupied) for $0\le\mu\le(15/2)V$, in addition to ``empty'' ($\mu\le 0$) and ``full'' ($\mu\ge(15/2)V$).

For $t>0$ the minimum eigenvalue of (\ref{3-44}) is
\small
\ba
\lambda_{\rm min}&=&E_0+30V\rho_{\rm A}+24V\rho_{\rm B}+36V\rho_{\rm C}-16\mu-6\sqrt{(2V\rho_{\rm B}+3V\rho_{\rm C}-\mu)^2+4t^2|2\phi_{\rm B}+3\phi_{\rm C}|^2}
\nonumber \\
&-&4\sqrt{(3V\rho_{\rm A}+3V\rho_{\rm C}-\mu)^2+36t^2|\phi_{\rm A}+\phi_{\rm C}|^2}
\nonumber \\
&-&6\sqrt{(3V\rho_{\rm A}+2V\rho_{\rm B}+V\rho_{\rm C}-\mu)^2+4t^2|3\phi_{\rm A}+2\phi_{\rm B}+\phi_{\rm C}|^2}\,.
\label{3-46}
\ea
\normalsize
Arguing similarly as done for the TH model, we are allowed to take $\phi_{\rm A},\phi_{\rm B}$, and $\phi_{\rm C}$ as real and positive. By making $\lambda_{\rm min}$ stationary, we eventually obtain six coupled equations for the six unknown parameters:
\ba
2\rho_{\rm B}+3\rho_{\rm C}&=&\frac{5}{2}-\frac{3V\rho_{\rm A}+3V\rho_{\rm C}-\mu}{\sqrt{\circled{\rm B}}}-\frac{3}{2}\frac{3V\rho_{\rm A}+2V\rho_{\rm B}+V\rho_{\rm C}-\mu}{\sqrt{\circled{\rm C}}}\,;
\nonumber \\
\rho_{\rm A}+\rho_{\rm C}&=&1-\frac{2V\rho_{\rm B}+3V\rho_{\rm C}-\mu}{2\sqrt{\circled{\rm A}}}-\frac{3V\rho_{\rm A}+2V\rho_{\rm B}+V\rho_{\rm C}-\mu}{2\sqrt{\circled{\rm C}}}\,;
\nonumber \\
\rho_{\rm A}+\rho_{\rm B}&=&1-\frac{2V\rho_{\rm B}+3V\rho_{\rm C}-\mu}{2\sqrt{\circled{\rm A}}}-\frac{3V\rho_{\rm A}+3V\rho_{\rm C}-\mu}{2\sqrt{\circled{\rm B}}}\,;
\nonumber \\
2\phi_{\rm B}+3\phi_{\rm C}&=&\frac{6t(\phi_{\rm A}+\phi_{\rm C})}{\sqrt{\circled{\rm B}}}+\frac{3t(3\phi_{\rm A}+2\phi_{\rm B}+\phi_{\rm C})}{\sqrt{\circled{\rm C}}}\,;
\nonumber \\
\phi_{\rm A}+\phi_{\rm C}&=&\frac{t(2\phi_{\rm B}+3\phi_{\rm C})}{\sqrt{\circled{\rm A}}}+\frac{t(3\phi_{\rm A}+2\phi_{\rm B}+\phi_{\rm C})}{\sqrt{\circled{\rm C}}}\,;
\nonumber \\
\phi_{\rm A}+\phi_{\rm B}&=&\frac{t(2\phi_{\rm B}+3\phi_{\rm C})}{\sqrt{\circled{\rm A}}}+\frac{3t(\phi_{\rm A}+\phi_{\rm C})}{\sqrt{\circled{\rm B}}}
\label{3-47}
\ea
with
\ba
\circled{\rm A}&=&(2V\rho_{\rm B}+3V\rho_{\rm C}-\mu)^2+4t^2(2\phi_{\rm B}+3\phi_{\rm C})^2\,;
\nonumber \\
\circled{\rm B}&=&(3V\rho_{\rm A}+3V\rho_{\rm C}-\mu)^2+36t^2(\phi_{\rm A}+\phi_{\rm C})^2\,;
\nonumber \\
\circled{\rm C}&=&(3V\rho_{\rm A}+2V\rho_{\rm B}+V\rho_{\rm C}-\mu)^2+4t^2(3\phi_{\rm A}+2\phi_{\rm B}+\phi_{\rm C})^2\,.
\label{3-48}
\ea

%
%
\begin{figure}
\begin{center}
\includegraphics[width=16cm]{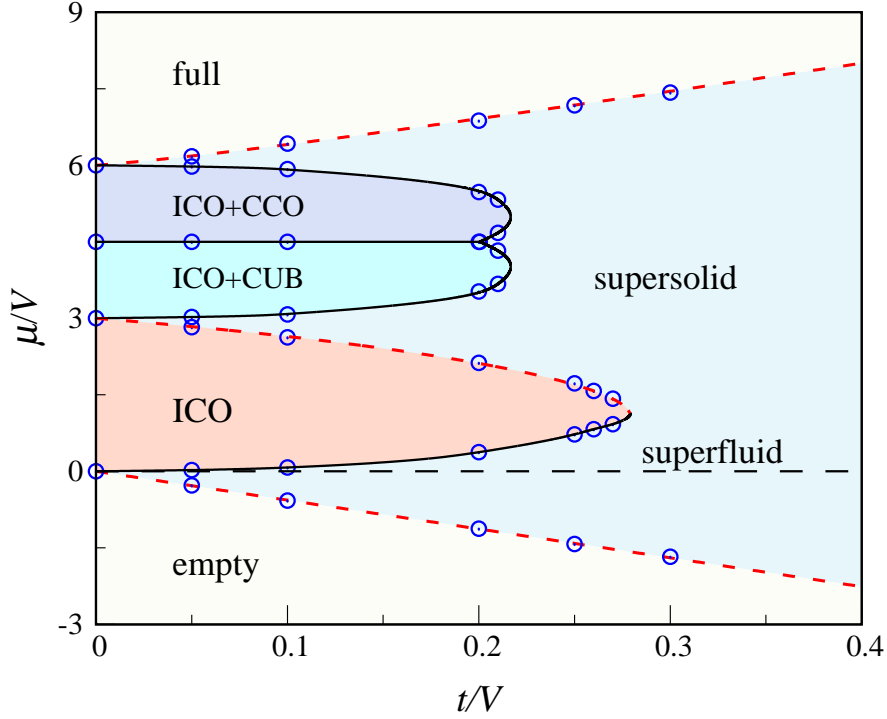}
\end{center}
\caption{MF phase diagram of the extended BH model with $U=+\infty$ on a PD graph, using $V$ as unit of energy. The blue dots mark transition points. The long-dashed $\mu=0$ line is where the system is superfluid. The dashed red curves are the continuous-transition loci derived in the text (cf. Eqs.~(\ref{3-49})). The remaining black lines represent first-order transitions.}
\label{fig8}
\end{figure}

The resulting phase diagram at $T=0$ is represented in Fig.\,8. We count as many as six distinct phases (seven, if we include the superfluid line $\mu=0$). Notice, in particular, how wide is the supersolid region, while the superfluid is confined to just a line. The insulating phases in Fig.\,8 are the same as found for $t=0$, and the ICO+CUB and ICO+CCO lobes are specular to each other with respect to $\mu=(9/2)V$. At variance with the TH model, where A+B and A+C phases are indistinguishable (i.e., degenerate), ICO+CUB and ICO+CCO are distinct phases, each with its own lobe in the phase diagram. The continuous-transition lines are three: those separating ``empty" and ``full'' from the supersolid region, and the descending part of the line between ICO and the supersolid. In the latter phase, the order parameters are symmetric between B and C, as implied by the data reported in Fig.\,9. Moreover, we see that $\rho_{\rm A}>\rho_{\rm B}=\rho_{\rm C}$ for $\mu>0$ and $\rho_{\rm A}\lesssim\rho_{\rm B}=\rho_{\rm C}$ for $\mu<0$.

%
%
\begin{figure}
\begin{center}
\includegraphics[width=8.1cm]{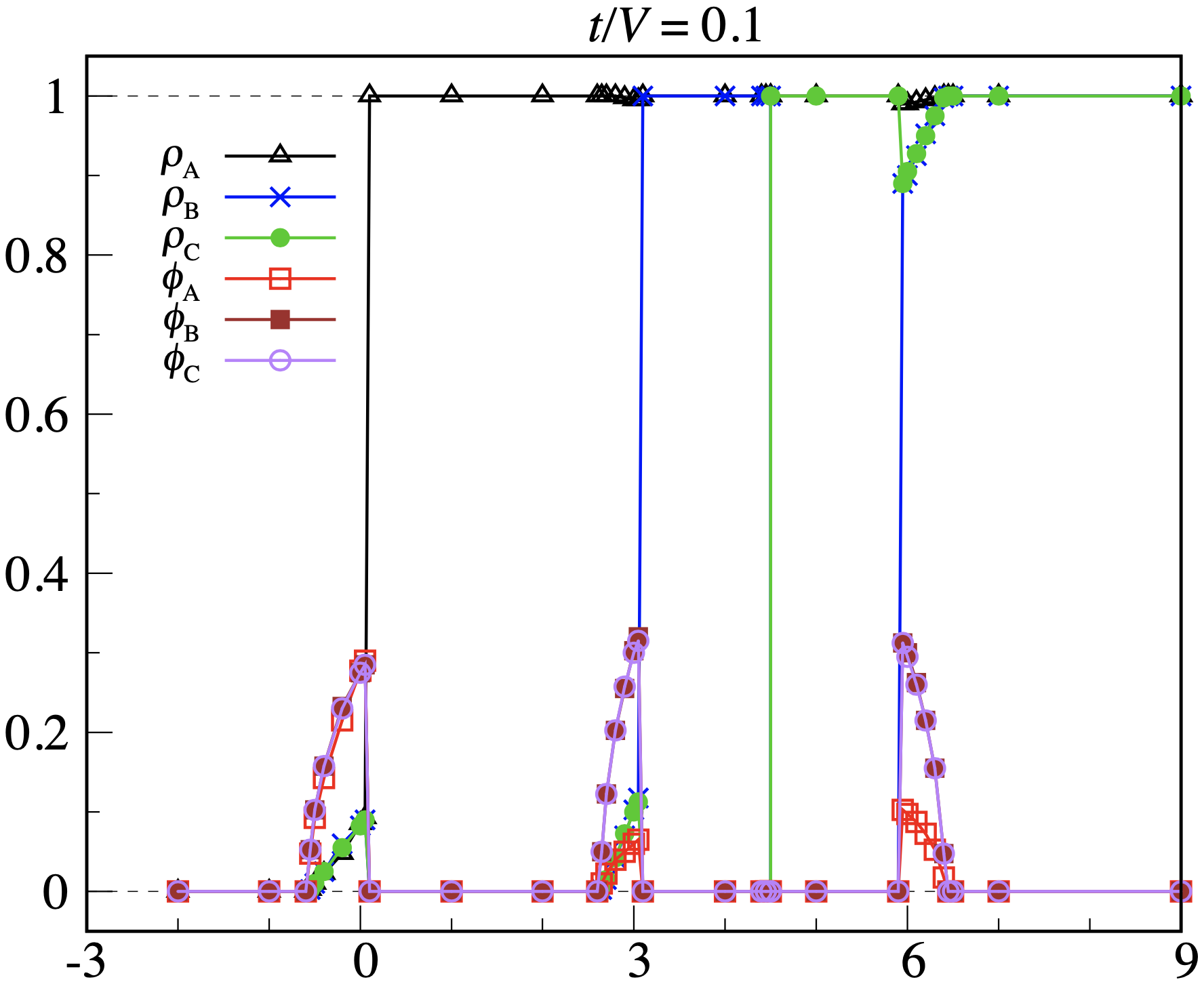}
\includegraphics[width=8.1cm]{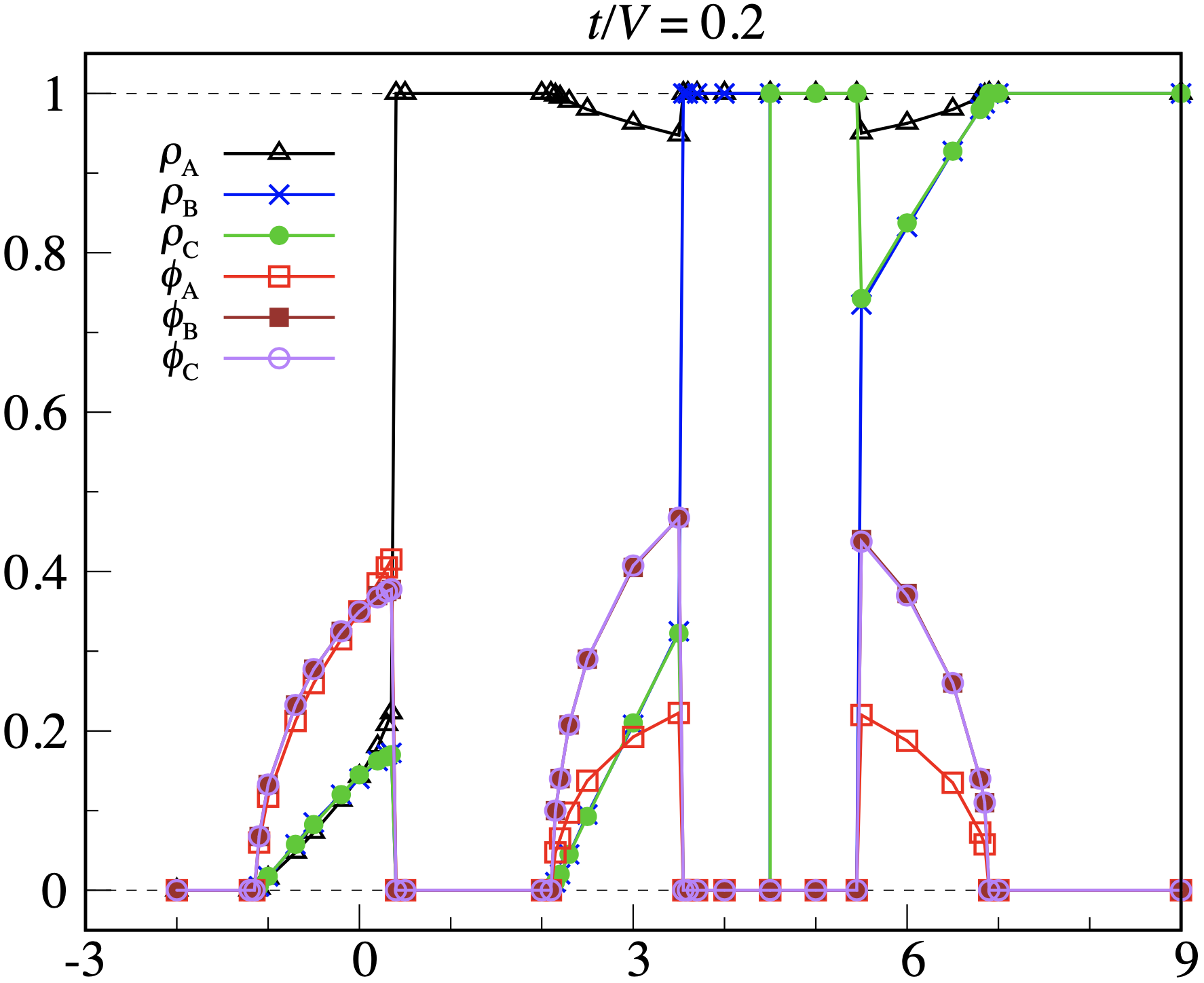}
\includegraphics[width=8.1cm]{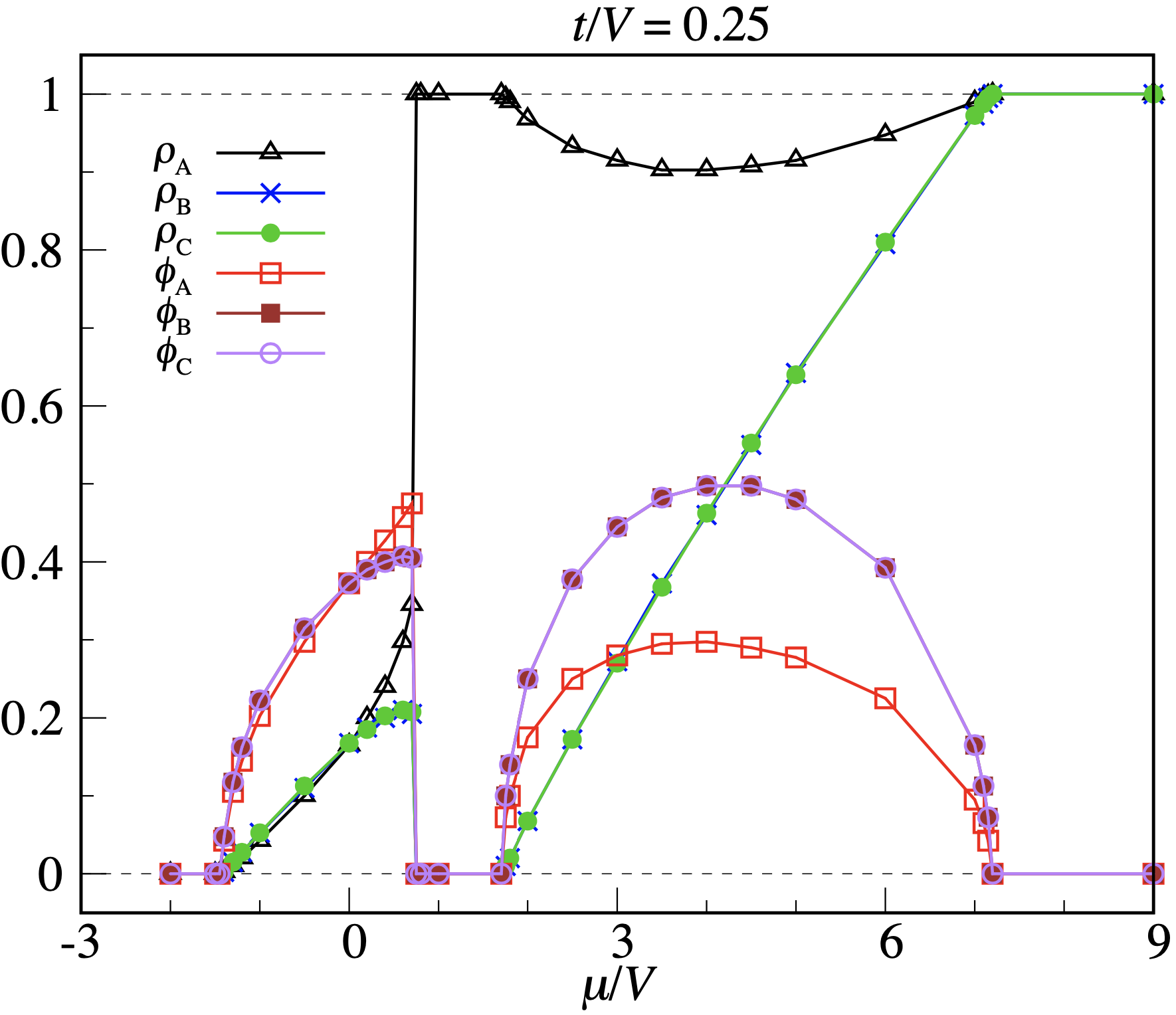}
\includegraphics[width=8.1cm]{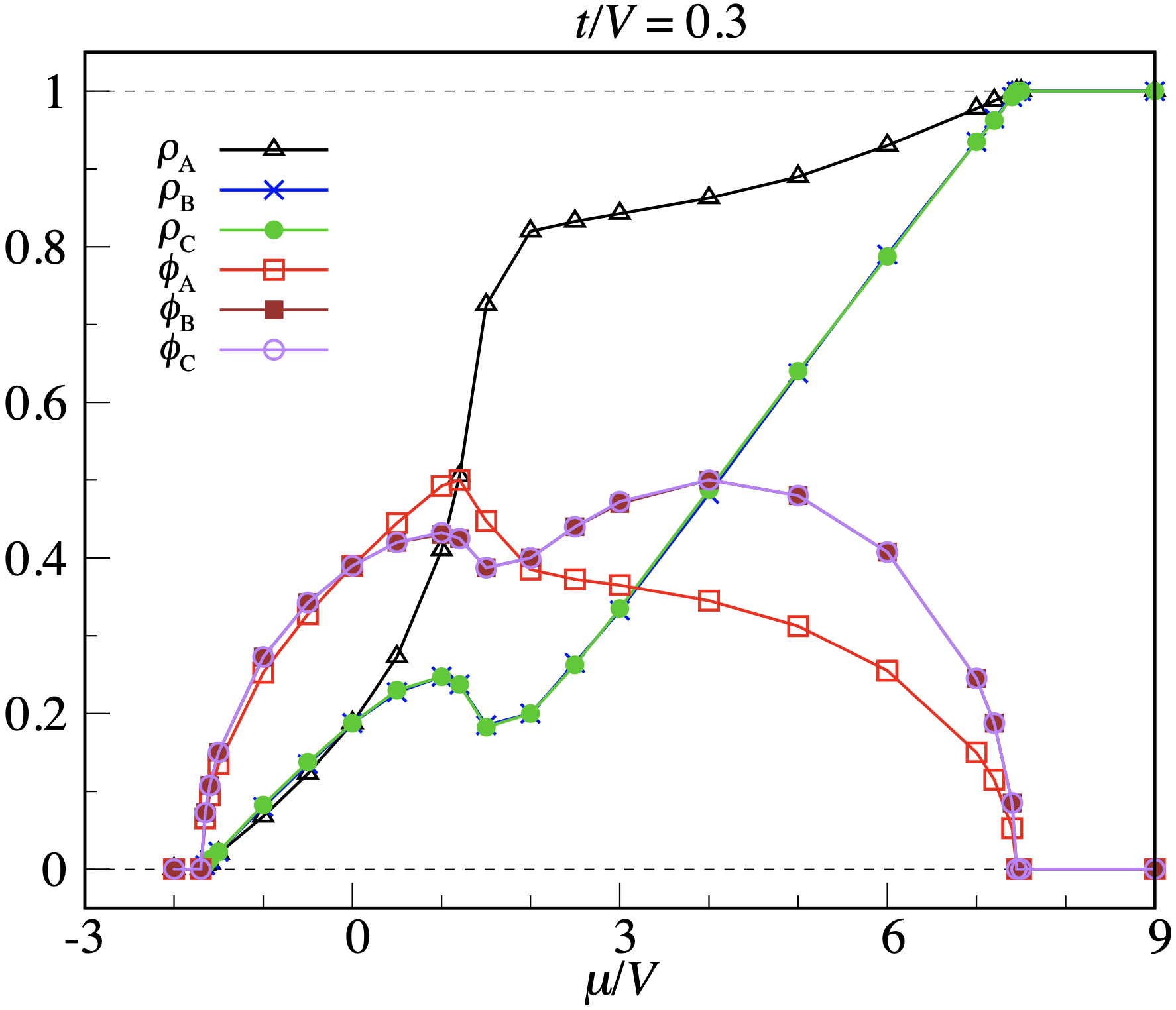}
\end{center}
\caption{Extended BH model on the PD graph. Order parameters plotted as a function of $\mu$ for fixed $t$ (from top left to bottom right, $t/V=0.1,0.2,0.25,0.3$).}
\label{fig9}
\end{figure}

Using B-C symmetry, we may simplify Eqs.\,(\ref{3-47}) and (\ref{3-48}) and determine the equations of the continuous-transition loci by following the same procedure used for the TH model. We find:
\ba
\mu&=&-\frac{3+\sqrt{69}}{2}t\qquad\qquad\qquad\qquad\qquad({\rm ``empty"\text{-}supersolid\,\,boundary})\,;
\nonumber \\
\mu&=&\frac{11V+3t+\sqrt{V^2+6Vt+69t^2}}{2}\qquad({\rm ``full"\text{-}supersolid\,\,boundary})\,;
\nonumber \\
\mu&=&\frac{3V-3t+\sqrt{9V^2-18Vt-51t^2}}{2}\qquad({\rm ICO\text{-}supersolid\,\,top\,\,boundary})\,.
\label{3-49}
\ea
In particular, upon requiring in the latter expression that $9V^2-18Vt-51t^2\ge 0$, the coordinates of the tricritical point are $t_c=(2\sqrt{15}-3)V/17=0.2791\ldots\,V$ and $\mu_c=(3V-3t_c)/2=1.0812\ldots\,V$.

\section{Conclusions}

The extended BH model is arguably the simplest model of quantum many-body system where one can accurately study, already in mean-field approximation, the onset of crystalline order and its interplay with superfluid order. Especially, this model provides a theoretical framework where supersolid phases, combining crystalline order with broken $U(1)$ symmetry, appear quite naturally and can thus be thoroughly examined.

In this paper the focus is on crystalline-like arrangements of spinless bosons placed on the nodes of a semiregular spherical mesh. We have considered two cases: the graph of a tetrakis hexahedron, where we find a ground state with octahedral symmetry; and the graph of a pentakis dodecahedron, where we find a ground state with icosahedral symmetry. Needless to say, ground states with polyhedral symmetry can only be stable for values of the hopping parameter $t$ that are small relative to the repulsion strength $V$. For larger $t$ values, wandering of particles throughout the nodes is no longer forbidden and the condensed fraction becomes non-zero. At variance with the extended BH model on a lattice, the presence in semiregular graphs of subsets of inequivalent nodes is at the origin of the destabilization of superfluidity towards supersolidity, which thus occurs in a wide region of thermodynamic parameters.

Clearly, no true phases or phase transitions can exist in a finite system, but only approximate orders with smooth crossovers between them. This weakness of our theory turns into an opportunity when we realize that mean-field theory can be checked against exact diagonalization. We have made this comparison for the smaller of our graphs (i.e., the skeleton of a tetrakis hexahedron), highlighting the many similarities and a few differences. Arrays of traps centered at the vertices of a polyhedron can now be realized and loaded with Rydberg atoms through moving optical tweezers~\cite{Barredo,Browaeys}, thus making it possible to check our predictions in systems of bosonic atoms.

\section{Acknowledgements}
I am grateful to an anonymous Referee for pointing out Refs.\,29, 39, and 59, allowing me to expand the scope of the paper.

\newpage
\appendix
\section{Partition function and thermal averages for a local Hamiltonian}
\renewcommand{\theequation}{A.\arabic{equation}}

In this Appendix we recall a few properties of a lattice boson Hamiltonian $H$ in which sites --- not particles --- are fully decoupled,
\be
H=\sum_{i=1}^Mh_i\,,
\label{a-1}
\ee
where $M$ is the number of lattice sites and, e.g., $h_1$ (a function of $a_1^\dagger$ and $a_1$) operates in the subspace ${\cal F}_1$ generated by $\va\equiv\ve{0,0,\ldots},\ve{1,0,\ldots},\ve{2,0,\ldots}$, and so on. For such a $H$, the eigenfunctions take the form of Gutzwiller~\cite{Rokhsar,Krauth},
\be
\ve\psi=\big(\prod_iG_i\big)\va\,\,\,\,\,\,{\rm with}\,\,\,\,\,\,G_i\equiv\sum_{n=0}^\infty c_n(i)\frac{\big(a_i^\dagger\big)^n}{\sqrt{n!}}\,,
\label{a-2}
\ee
provided that $G_i\va$ is eigenfunction of $h_i$:
\be
h_iG_i\va=\epsilon_iG_i\va\,.
\label{a-3}
\ee
Indeed, since operators at different sites commute, for $i=1$
\be
h_1\ve\psi=h_1G_1G_2\cdots G_M\va=G_2\cdots G_Mh_1G_1\va=\epsilon_1G_2\cdots G_MG_1\va=\epsilon_1\ve\psi\,,
\label{a-4}
\ee
and similarly for the other sites, implying
\be
H\ve\psi=\sum_i\epsilon_i\ve\psi\,.
\label{a-5}
\ee
The Fock states $\ve{n_1,n_2,\ldots}$ are Gutzwiller states where only one coefficient $c_n(i)$ is non-zero for each $i$, but they are usually not energy eigenstates. In the following we assume $\sum_n\big|c_n(i)\big|^2=1$ for $i=1,\ldots,M$, in such a way that $\me\psi\psi=1$. In terms of Fock states, the eigenfunction (\ref{a-2}) is written as
\be
\ve\psi=\sum_{n_1,\ldots,n_M}c_{n_1}(1)\cdots c_{n_M}(M)\ve{n_1,\ldots,n_M}\,.
\label{a-6}
\ee
It is worth emphasizing the factorized structure exhibited by the Fourier coefficients, which is an effect of the strictly local nature of the Hamiltonian (\ref{a-1}).

Applying the basic rules of creation and annihilation operators, it follows for every $i$ and $\ve\psi$ of type (\ref{a-6}) that
\be
\me\psi{a_i\psi}=\meb{a_i^\dagger\psi}\psi=\sum_{n=0}^\infty\sqrt{n+1}\,c_n^*(i)c_{n+1}(i)\equiv\psi(i)\,\,\,\,\,\,{\rm and}\,\,\,\,\,\,\meb\psi{a_i^\dagger a_i\psi}=\sum_{n=1}^\infty n|c_n(i)|^2\,.
\label{a-7}
\ee
Moreover, the average of $a_i^\dagger a_j$ for $i\ne j$ is factorized:
\be
\meb\psi{a_i^\dagger a_j\psi}=\psi^*(i)\psi(j)=\meb\psi{a_i^\dagger\psi}\me\psi{a_j\psi}\,,
\label{a-8}
\ee
which holds in particular for $\ve\psi$ being the ground state of $H$.

To calculate the thermal average of, say, $a_1^\dagger a_2$ we need a complete set of energy eigenfunctions. To this aim, we first diagonalize each $h_i$ in its domain ${\cal F}_i$ (in practice, some cutoff $n_{\rm max}$ is put on $n$ to account for the fact that large $n$ values are energetically suppressed). We denote $\big\{\veb{\psi^{(\alpha)}}=G_1^{(\alpha_1)}\cdots G_M^{(\alpha_M)}\va,\alpha_i=1,2,\ldots,n_{\rm max}\big\}$ a complete set of orthonormal eigenfunctions of $H$ (observe that the total number of eigenfunctions is $n_{\rm max}^M$, same as the number of Fock states $\ve{n_1,\ldots,n_M}$). Then, the partition function reads:
\ba
Z&=&{\rm Tr}\big(e^{-\beta H}\big)=\sum_{\alpha_1,\ldots,\alpha_M}\me{\psi^{(\alpha)}}{e^{-\beta H}\psi^{(\alpha)}}=\sum_{\alpha_1,\ldots,\alpha_M}e^{-\beta\big(\epsilon_1^{(\alpha_1)}+\ldots+\epsilon_M^{(\alpha_M)}\big)}
\nonumber \\
&=&\sum_{\alpha_1}e^{-\beta\epsilon_1^{(\alpha_1)}}\cdots\sum_{\alpha_M}e^{-\beta\epsilon_M^{(\alpha_M)}}\,.
\label{a-9}
\ea
Since each eigenfunction can be expanded on the Fock basis as in Eq.~(\ref{a-6}), we have
\be
\meb{\psi^{(\alpha)}}{a_1^\dagger a_2\psi^{(\alpha)}}=\psi^*(1,\alpha_1)\psi(2,\alpha_2)\,,
\label{a-10}
\ee
where, for example, $\psi(1,\alpha_1)=\sum_{n=0}^\infty\sqrt{n+1}\,c_n^*(1,\alpha_1)c_{n+1}(1,\alpha_1)$. In the end, we find:
\ba
\big<a_1^\dagger a_2\big>&=&\frac{1}{Z}{\rm Tr}\big(e^{-\beta H}a_1^\dagger a_2\big)=\frac{1}{Z}\sum_{\alpha_1}e^{-\beta\epsilon_1^{(\alpha_1)}}\psi^*(1,\alpha_1)\sum_{\alpha_2}e^{-\beta\epsilon_2^{(\alpha_2)}}\psi(2,\alpha_2)\sum_{\alpha_3}e^{-\beta\epsilon_3^{(\alpha_3)}}\cdots
\nonumber \\
&=&\frac{\sum_{\alpha_1}e^{-\beta\epsilon_1^{(\alpha_1)}}\psi^*(1,\alpha_1)}{\sum_{\alpha_1}e^{-\beta\epsilon_1^{(\alpha_1)}}}\frac{\sum_{\alpha_2}e^{-\beta\epsilon_2^{(\alpha_2)}}\psi(2,\alpha_2)}{\sum_{\alpha_2}e^{-\beta\epsilon_2^{(\alpha_2)}}}=\big<a_1^\dagger\big>\langle a_2\rangle\,,
\label{a-11}
\ea
meaning that $a_i^\dagger$ and $a_j$ are uncorrelated not only at $T=0$ but for all temperatures. One may similarly show that $\big<n_in_j\big>=\big<n_i\big>\big<n_j\big>$ for $i\ne j$.

If no external field is present, then the system is homogeneous and it is sufficient to diagonalize $h$ at one site only. In particular, the ground-state energy per site is simply the minimum eigenvalue of a $(n_{\rm max}+1)\times(n_{\rm max}+1)$ Hermitian matrix. However, if the lattice is bipartite (i.e., it consists of two disjoint sublattices, A and B, such that nearest-neighbor sites belong to different sublattices), then, depending on the Hamiltonian and on the control parameters, the ground state may also reflect the same checkerboard structure --- as occurs, for instance, in an extended BH model with nearest-neighbor repulsion, where the minimum-energy state may be a density wave or a supersolid state. In this case, the minimum energy is $M_{\rm A}\epsilon_{{\rm A},{\rm min}}+M_{\rm B}\epsilon_{{\rm B},{\rm min}}$, with sublattice energies obtained from the diagonalization of two distinct $(n_{\rm max}+1)\times(n_{\rm max}+1)$ matrices. Alternatively, we may view the system as a two-site BH model and represent the Hamiltonian on a basis of pair states, $\{\ve{n_{\rm A},n_{\rm B}}\}$, as done in Refs.~\cite{Gheeraert,Prestipino6}.

\section{Variational foundation of the DA}
\renewcommand{\theequation}{B.\arabic{equation}}

We show hereafter that the DA treatment of the extended BH model may be justified as an application of the variational method based on the Gibbs-Bogoliubov (GB) inequality, also valid for a quantum system~\cite{Carlen}. Hence, the self-consistent DA parameters are also those parameters that ensure minimization of a variational grand potential, as is usual in classical and quantum phase-diagram reconstruction (see examples in Refs.~\cite{Prestipino7,Prestipino8,Prestipino9,Kunimi,Prestipino10}).

Let the extended BH Hamiltonian be written as
\be
H=-t\sum_{ij}z_{ij}a_i^\dagger a_j+\frac{V}{2}\sum_{ij}z_{ij}n_in_j+\sum_if(n_i)\,,
\label{b-1}
\ee
where $z_{ij}=1$ if $i$ and $j$ are NN sites and zero otherwise ($z_{ij}$ and its inverse are symmetric matrices). All local terms in the BH Hamiltonian, including the chemical-potential term, have been absorbed in $f(n_i)$. With the aim to estimate the grand potential $\Omega$ of (\ref{b-1}), we introduce a fully local Hamiltonian
\be
H_0=-t\sum_i\big(F_ia_i^\dagger+F_i^*a_i\big)+V\sum_iR_in_i+\sum_if(n_i)\,,
\label{b-2}
\ee
where $F_i\in\mathbb{C}$ and $R_i\in\mathbb{R}$ are parameters to be optimized. According to the GB inequality,
\be
\Omega\le\Omega_0+\langle H-H_0\rangle_0\equiv\Omega_{\rm GB}\,,
\label{b-3}
\ee
where $\langle\cdots\rangle_0$ is a thermal average over the Boltzmann distribution pertaining to $H_0$ and
\be
\Omega_0=-\frac{1}{\beta}\ln{\rm Tr}\,e^{\beta t\sum_i\left(F_ia_i^\dagger+F_i^*a_i\right)-\beta V\sum_iR_in_i-\beta\sum_if(n_i)}
\label{b-4}
\ee
is the grand potential of the trial Hamiltonian. Using equalities like (\ref{a-11}), we obtain:
\ba
\langle H-H_0\rangle_0&=&-t\sum_{ij}z_{ij}\langle a_i\rangle_0^*\langle a_j\rangle_0+t\sum_i\big(F_i\langle a_i\rangle_0^*+F_i^*\langle a_i\rangle_0\big)
\nonumber \\
&+&\frac{V}{2}\sum_{ij}z_{ij}\langle n_i\rangle_0\langle n_j\rangle_0-V\sum_iR_i\langle n_i\rangle_0\,.
\label{b-5}
\ea
The best parameters are those providing the absolute minimum of $\Omega_{\rm BG}$. As long as this minimum falls in the interior of parameter space, a necessary condition for it is the vanishing of first-order derivatives,
\be
\frac{\partial\Omega_{\rm BG}}{\partial F_k^*}=\left(\frac{\partial\Omega_{\rm BG}}{\partial F_k}\right)^*=0\,\,\,\,\,\,{\rm and}\,\,\,\,\,\,\frac{\partial\Omega_{\rm BG}}{\partial R_k}=0\,.
\label{b-6}
\ee
The former derivative is a conjugate cogradient, or Wirtinger derivative, and is to be interpreted as a partial derivative with respect to $F_k^*$, while keeping $F_k$ constant.

Before proceeding to the solution of Eqs.~(\ref{b-6}) we need another piece of information, since $F_i^*$ and $R_i$ enter in an intricate manner inside $\Omega_0$, see Eq.~(\ref{b-4}). Consider a Hamiltonian $\xi A+B$ where $\xi$ is a real or complex parameter and $A$ and $B$ are quantum observables independent of $\xi$. For such a Hamiltonian, the normalized eigenstates $\ve s$, such that $(\xi A+B)\ve s=E_s\ve s$, form a complete set. Then, the partition function reads
\be
Z(\xi)={\rm Tr}\,e^{-\beta(\xi A+B)}=\sum_s\meb{s}{e^{-\beta(\xi A+B)}s}=\sum_se^{-\beta E_s}\,.
\label{b-7}
\ee
By noting that (the components of) $\ve s$ and $E_s$ are both dependent on $\xi$, we obtain:
\be
\frac{\partial\ln Z}{\partial\xi}=-\frac{\beta}{Z}\sum_s\frac{\partial E_s}{\partial\xi}e^{-\beta E_s}
\label{b-8}
\ee
with
\ba
\frac{\partial E_s}{\partial\xi}&=&\meb{\partial_\xi s}{(\xi A+B)s}+\me s{As}+\meb s{(\xi A+B)\partial_\xi s}
\nonumber \\
&=& E_s\left[\me{\partial_\xi s}s+\me{\partial_\xi s}s^*\right]+\me s{As}
\nonumber \\
&=& E_s\partial_\xi\me ss+\me s{As}=\me s{As}\,,
\label{b-9}
\ea
in such a way that
\be
\frac{\partial\ln Z}{\partial\xi}=-\frac{\beta}{Z}\sum_s\me s{As}e^{-\beta E_s}=-\frac{\beta}{Z}{\rm Tr}\big(Ae^{-\beta(\xi A+B)}\big)=-\beta\langle A\rangle\,.
\label{b-10}
\ee

With the above result established, by simple algebra we obtain:
\ba
\frac{\partial\Omega_{\rm BG}}{\partial F_k^*}&=&-t\sum_i\left[\left(F_i-\sum_jz_{ij}\langle a_j\rangle_0\right)\frac{\partial\langle a_i\rangle_0^*}{\partial F_k^*}+\left(F_i^*-\sum_jz_{ij}\langle a_j\rangle_0^*\right)\frac{\partial\langle a_i\rangle_0}{\partial F_k^*}\right]
\nonumber \\
&&-V\sum_i\left(R_i-\sum_jz_{ij}\langle n_j\rangle_0\right)\frac{\partial\langle n_i\rangle_0}{\partial F_k^*}
\label{b-11}
\ea
and
\ba
\frac{\partial\Omega_{\rm BG}}{\partial R_k}&=&-t\sum_i\left[\left(F_i-\sum_jz_{ij}\langle a_j\rangle_0\right)\frac{\partial\langle a_i\rangle_0^*}{\partial R_k}+\left(F_i^*-\sum_jz_{ij}\langle a_j\rangle_0^*\right)\frac{\partial\langle a_i\rangle_0}{\partial R_k}\right]
\nonumber \\
&&-V\sum_i\left(R_i-\sum_jz_{ij}\langle n_j\rangle_0\right)\frac{\partial\langle n_i\rangle_0}{\partial R_k}\,.
\label{b-12}
\ea
In order that (\ref{b-11}) and (\ref{b-12}) be zero, it is sufficient (and seemingly also necessary) that
\be
F_i=\sum_jz_{ij}\langle a_j\rangle_0\,\,\,\,\,\,{\rm and}\,\,\,\,\,\,R_i=\sum_jz_{ij}\langle n_j\rangle_0\,.
\label{b-13}
\ee
Observe that the above equations define $F_i$ and $R_i$ only implicitly, since $\langle a_j\rangle_0$ and $\langle n_j\rangle_0$ are themselves dependent on these parameters. Upon formally inverting Eqs.~(\ref{b-13}) we find the equivalent relations
\be
\langle a_i\rangle_0=\sum_j\big(z^{-1}\big)_{ij}F_j\,\,\,\,\,\,{\rm and}\,\,\,\,\,\,\langle n_i\rangle_0=\sum_j\big(z^{-1}\big)_{ij}R_j\,.
\label{b-14}
\ee
The point of absolute minimum for $\Omega_{\rm BG}$ is among the solutions to Eqs.~(\ref{b-14}).

We now introduce another functional, $\widetilde\Omega_{\rm BG}=\Omega_0+\widetilde{\langle H-H_0\rangle_0}$, which is obtained from $\Omega_{\rm BG}$ by substituting the averages (\ref{b-14}) into (\ref{b-5}):
\be
\widetilde{\langle H-H_0\rangle_0}=t\sum_{ik}\big(z^{-1}\big)_{ik}F_iF_k^*-\frac{V}{2}\sum_{ik}\big(z^{-1}\big)_{ik}R_iR_k\,.
\label{b-15}
\ee
The new functional $\widetilde\Omega_{\rm BG}$ is {\em different} from $\Omega_{\rm BG}$, but they share the same stationary points {\em and} stationary values: indeed, it is easy to see that Eqs.~(\ref{b-14}) are still necessary and sufficient conditions for
\be
\frac{\partial\widetilde\Omega_{\rm BG}}{\partial F_k^*}=0\,\,\,\,\,\,{\rm and}\,\,\,\,\,\,\frac{\partial\widetilde\Omega_{\rm BG}}{\partial R_k}=0\,.
\label{b-16}
\ee
We stress, however, that the nature of extremal points may not be preserved in the transition from $\Omega_{\rm BG}$ to $\widetilde\Omega_{\rm BG}$, as second-order derivatives in these points are generally different for the two functionals. Using the shorthands
\be
\phi_i=\sum_j\big(z^{-1}\big)_{ij}F_j\,\,\,\,\,\,{\rm and}\,\,\,\,\,\,\rho_i=\sum_j\big(z^{-1}\big)_{ij}R_j\,,
\label{b-17}
\ee
we may also write
\be
\widetilde\Omega_{\rm BG}=-\frac{1}{\beta}\ln{\rm Tr}\,e^{\beta t\sum_i\big(F_ia_i^\dagger+F_i^*a_i-F_i\phi_i^*\big)-\beta\frac{V}{2}\sum_i(2R_in_i-R_i\rho_i)-\beta\sum_if(n_i)}\,,
\label{b-18}
\ee
showing that $\widetilde\Omega_{\rm BG}$ is the grand potential of the DA Hamiltonian (\ref{2-3}). The values of $F_i$ and $R_i$ must then be selected imposing the (\ref{b-16}) or, equivalently, the (\ref{b-14}). If more solutions are found, we must choose the one that provides the minimum $\widetilde\Omega_{\rm BG}$ for the given $t,\mu$, and $T$.

\section{Derivation of the DA from the Hubbard-Stratonovich formula}
\renewcommand{\theequation}{C.\arabic{equation}}

The DA may also be justified using the language of functional integrals, as shown in Refs.~\cite{Fisher,Sheshadri,vanOosten} for the original BH model. We hereafter retrace the steps of this derivation making now reference to the extended BH model.

In the coherent-state representation, the partition function of a bosonic lattice Hamiltonian in normal-ordered form can be written as an integral over $M$ (i.e., as many as are the lattice sites) closed paths. For the extended BH model, in the continuum limit one finds:
\ba
\Xi&=&\oint\prod_k{\cal D}\phi_k{\cal D}\phi_k^*\,e^{-\hbar^{-1}S[\phi,\phi^*]}\,\,\,\,\,\,{\rm with}
\nonumber \\
S[\phi,\phi^*]&=&\int_0^{\beta\hbar}{\rm d}\tau\bigg[\underbrace{\sum_i\phi_i^*(\hbar\partial_\tau-\mu)\phi_i+\frac{U}{2}\sum_i|\phi_i|^4}_{H^{(1)}(\phi^*,\phi)}-t\sum_{ij}z_{ij}\phi_i^*\phi_j+\frac{V}{2}\sum_{ij}z_{ij}|\phi_i|^2|\phi_j|^2\bigg]\,.
\nonumber \\
\label{c-1}
\ea
In the above formula, $\tau$ is the imaginary time and $S$ is the Euclidean action --- a functional of $M$ complex fields $\phi_i(\tau)$ and their conjugate fields $\phi_i^*(\tau)$, only subject to $\phi_i(0)=\phi_i(\beta\hbar)$. Furthermore, $H^{(1)}(\phi^*,\phi)$ is the symbol of the on-site terms in the Hamiltonian. Compared to the operator formalism, the coherent-state path integral offers the distinct advantage that any complications due to non-commuting observables are swept away ($\phi_i(\tau)$ is an ordinary, albeit complex, function of a real variable). The price to pay is the introduction of an extra time variable and of the ubiquitous $\phi^*\partial_\tau\phi$ term in the action.

The idea behind the application of the Hubbard-Stratonovich (HS) formula is to decouple the interaction terms in (\ref{c-1}) by employing a suitable integral identity, even though at the price of introducing more fields. In particular, we will need a (dimensionless) complex field $M_i$ for the hopping term and a (dimensionless) real field $N_i$ for the term proportional to $V$, for each $i$. The HS formula is just another name for the Gaussian integral; for a complex matrix $A$ with a positive-definite Hermitian part, it reads:
\be
{\cal N}\int\prod_{k=1}^M{\rm d}\Re z_k{\rm d}\Im z_k^*\,e^{-\sum_{ij}z_i^*A_{ij}z_j}=1\,\,\,\,\,\,{\rm with}\,\,\,\,\,\,{\cal N}=\frac{\det A}{\pi^M}\,.
\label{c-2}
\ee
By resorting to the identities
\ba
&&-\sum_{ij}\big(M_i^*-\sum_mz_{im}\phi_m^*\big)(z^{-1})_{ij}\big(M_j-\sum_nz_{jn}\phi_n\big)
\nonumber \\
&=&-\sum_{ij}(z^{-1})_{ij}M_i^*M_j+\sum_i(M_i^*\phi_i+M_i\phi_i^*)-\sum_{ij}z_{ij}\phi_i^*\phi_j
\label{c-3}
\ea
and
\ba
&&-\frac{1}{2}\sum_{ij}z_{ij}|\phi_i|^2|\phi_j|^2=\frac{1}{2}\sum_{ij}(z^{-1})_{ij}N_iN_j-\sum_iN_i|\phi_i|^2
\nonumber \\
&-&\frac{1}{2}\sum_{ij}\big(N_i-\sum_mz_{im}|\phi_m|^2\big)(z^{-1})_{ij}\big(N_j-\sum_nz_{jn}|\phi_n|^2\big)\,,
\label{c-4}
\ea
the partition function (\ref{c-1}) can be rewritten as
\be
\Xi=\int\prod_k{\cal D}M_k{\cal D}M_k^*{\cal D}N_k\,e^{-\hbar^{-1}S_{\rm eff}[M,M^*,N]}
\label{c-5}
\ee
with
\footnotesize
\ba
&&S_{\rm eff}=\int_0^{\beta\hbar}{\rm d}\tau\,t\sum_{ij}(z^{-1})_{ij}M_i^*(\tau)M_j(\tau)-\int_0^{\beta\hbar}{\rm d}\tau\,\frac{V}{2}\sum_{ij}(z^{-1})_{ij}N_i(\tau)N_j(\tau)
\nonumber \\
&-&\hbar\ln\oint\prod_k{\cal D}\phi_k{\cal D}\phi_k^*\,\exp\left\{-\hbar^{-1}\underbrace{\int_0^{\beta\hbar}{\rm d}\tau\left[H^{(1)}(\phi^*,\phi)-t\sum_i\big(M_i^*\phi_i+M_i\phi_i^*\big)+V\sum_iN_i|\phi_i|^2\right]}_{S_{\rm DA}}\right\}\,.
\nonumber \\
\label{c-6}
\ea
\normalsize
The normalization factors arising from the Gaussian integrals have been absorbed in the integration measure. We note the formal similarity between the effective action (\ref{c-6}) and the functional $\widetilde\Omega_{\rm BG}$ in Eq.~(\ref{b-18}).

As for the partition function (\ref{c-5}), a natural MF estimate is obtained by approximating it with the integrand evaluated at the saddle point. The ``coordinates'' of the saddle point are determined through the equations
\ba
0=\frac{\delta S_{\rm eff}}{\delta M_i^*(\tau_1)}&=&t\sum_j(z^{-1})_{ij}M_j(\tau_1)-t\frac{\int\prod_k{\cal D}\phi_k{\cal D}\phi_k^*\,\phi_i(\tau_1)e^{-\hbar^{-1}S_{DA}}}{\int\prod_k{\cal D}\phi_k{\cal D}\phi_k^*\,e^{-\hbar^{-1}S_{DA}}}
\nonumber \\
&\Longrightarrow&\,\,\,\,\,\,M_i=\sum_jz_{ij}\langle\phi_j(\tau_1)\rangle_{\rm DA}
\label{c-7}
\ea
and
\ba
0=\frac{\delta S_{\rm eff}}{\delta N_i(\tau_1)}&=&-V\sum_j(z^{-1})_{ij}N_j(\tau_1)+V\frac{\int\prod_k{\cal D}\phi_k{\cal D}\phi_k^*\,|\phi_i(\tau_1)|^2e^{-\hbar^{-1}S_{DA}}}{\int\prod_k{\cal D}\phi_k{\cal D}\phi_k^*\,e^{-\hbar^{-1}S_{DA}}}
\nonumber \\
&\Longrightarrow&\,\,\,\,\,\,N_i=\sum_jz_{ij}\langle|\phi_j(\tau_1)|^2\rangle_{\rm DA}\,.
\label{c-8}
\ea
Clearly, Eqs.~(\ref{c-7}) and (\ref{c-8}) are analogous to Eqs.~(\ref{b-13}) above.

\section{Mean-field treatment of hard-core bosons in spin language}
\renewcommand{\theequation}{D.\arabic{equation}}

We originally owe to Matsubara and Matsuda~\cite{Matsubara} the observation that a second-quantized Hamiltonian for {\em hard-core bosons} can be rephrased in terms of half-unit spins:
\be
a_i^\dagger=S_i^+\equiv S_i^x+iS_i^y\,\,\,\,\,\,\big({\rm hence}\,\,\,a_i=S_i^-\equiv S_i^x-iS_i^y\,\,\,{\rm and}\,\,\,n_i=S_i^z+1/2\big)\,.
\label{d-1}
\ee
Thus, an occupied site is represented by an up spin, while an empty site is represented by a down spin. This mapping has been exploited in many studies of the BH model (see, e.g., Refs.~\cite{Bruder,Scalettar,Murthy}). For hard-core bosons, creation and annihilation operators at different sites commute, while $a_i$ and $a_i^\dagger$ are {\em anticommuting} operators as a result of the dynamical suppression of Fock states with two or more particles per site (see, e.g., \cite{Morita}).

For the extended BH model with infinite $U$ the equivalent spin Hamiltonian is readily found to be
\be
H_S=-J_\perp\sum_{\langle i,j\rangle}\big(S_i^xS_j^x+S_i^yS_j^y\big)+J_\parallel\sum_{\langle i,j\rangle}S_i^zS_j^z-H_z\sum_iS_i^z+C\,,
\label{d-2}
\ee
where $J_\perp=2t$ is a ferromagnetic transverse exchange, $J_\parallel=V$ is an antiferromagnetic longitudinal exchange, $H_z=\mu-zV/2$ ($z$ being the lattice coordination number) is an external magnetic field, and $C=MzV/8-M\mu/2$ is an offset. The Hamiltonian (\ref{d-2}) is a spin-$1/2$ $XXZ$ Heisenberg model. Had we adopted the different convention of Matsuda and Tsuneto~\cite{Matsuda}, that is $a_i^\dagger=S_i^-$, we would have got the same Hamiltonian as in (\ref{d-2}) but for the sign in front of the magnetization term. A modulated density of the original BH system corresponds to finite wavevector Ising-type order of the spins. Similarly, superfluidity maps to ferromagnetic spin ordering in the $x$-$y$ plane. In units of $J_\parallel=V$, the spin Hamiltonian reads
\be
H_S=\sum_{\langle i,j\rangle}\left[S_i^zS_j^z-\Delta\big(S_i^xS_j^x+S_i^yS_j^y\big)\right]-h\sum_iS_i^z+C/V
\label{d-3}
\ee
with $\Delta=2t/V$ and $h=\mu/V-z/2$. Spin systems like the one described by $H_S$ can actually be studied with ultracold Rydberg atoms~\cite{Signoles,Browaeys}, which would allow to observe the ground states of our hard-core boson model in a real system.

In MF theory, the spins are treated as they were classical: ${\bf S}_i=(S_i^x,S_i^y,S_i^z)$ is an ordinary vector of magnitude $S=1/2$ for every $i$. For $T=0$, the problem is then reduced to mapping the spin configuration of minimum energy as a function of $t$ and $\mu$. For the Hamiltonian (\ref{d-3}), which is rotationally symmetric in the $x$-$y$ plane, we may assume that all spins lie in the $x$-$z$ plane. Putting ${\bf S}_i=(1/2)\mathbf{\Omega}_i$, the MF Hamiltonian reads (neglecting the unnecessary $C/V$ constant):
\be
H_{\rm MF}=\frac{1}{4}\sum_{\langle i,j\rangle}\big(\Omega_i^z\Omega_j^z-\Delta\,\Omega_i^x\Omega_j^x\big)-\frac{h}{2}\sum_i\Omega_i^z\,.
\label{d-4}
\ee

As a matter of example, let us reconsider the QCT model of hard-core bosons on the vertices of a cube ($M=8,z=3$)~\cite{Prestipino6}. Due to the bipartite structure of the lattice, the MF energy $E_S$ can be parametrized in terms of the orientation of two unit vectors only, $\mathbf{\Omega}_A$ and $\mathbf{\Omega}_B$, in the assumption that spins are identical on the sites of the same sublattice:
\be
E_S=3(\cos\theta_A\cos\theta_B-\Delta\sin\theta_A\sin\theta_B)-2h(\cos\theta_A+\cos\theta_B)\,,
\label{d-5}
\ee
where $\theta_A$ ($\theta_B$) is the angle made by $\mathbf{\Omega}_A$ ($\mathbf{\Omega}_B$) with the positive $z$ axis. For $h=0$, which is tantamount to $\mu=(3/2)V$, the task of minimizing $E_S$ is easily accomplished:
\ba
&&{\rm if}\,\,\,\Delta>1\,\,\,{\rm then}\,\,\,\theta_A=\theta_B=\frac{\pi}{2}\,\,\,\longrightarrow\,\,\,{\rm superfluid}\,\,\,\,\,\,\big(\rightrightarrows\,,\,\leftleftarrows\big)\,;
\nonumber \\
&&{\rm if}\,\,\,\Delta<1\,\,\,{\rm then}\,\,\,\theta_A=0,\pi\,\,\,{\rm or}\,\,\,\theta_B=\pi,0\,\,\,\longrightarrow\,\,\,{\rm N\acute{e}el\,\,order}\,\,\,\,\,\,\big(\uparrow\downarrow\,,\,\downarrow\uparrow\big)\,.
\label{d-6}
\ea
Moreover, it is clear that for $h\gg 0$ the minimum of $E_S$ is attained for $\theta_A=\theta_B=0$ ($\uparrow\uparrow$), while for $h\ll 0$ the minimum falls at $\theta_A=\theta_B=\pi$ ($\downarrow\downarrow$). The analysis is simple also for $\Delta=0$ ($t=0$):
\ba
&&{\rm if}\,\,\,h<-\frac{3}{2}\,\,\,{\rm then}\,\,\,\theta_A=\theta_B=\pi\,\,\,{\rm and}\,\,\,E_S=3+4h\,\,\,\,\,\,\big(\downarrow\downarrow\big)\,;
\nonumber \\
&&{\rm if}\,\,\,-\frac{3}{2}<h<\frac{3}{2}\,\,\,{\rm then}\,\,\,\theta_A=0,\pi\,\,\,{\rm or}\,\,\,\theta_B=\pi,0\,\,\,{\rm and}\,\,\,E_S=-3\,\,\,\,\,\,\big(\uparrow\downarrow\,,\,\downarrow\uparrow\big)\,;
\nonumber \\
&&{\rm if}\,\,\,h>\frac{3}{2}\,\,\,{\rm then}\,\,\,\theta_A=\theta_B=0\,\,\,{\rm and}\,\,\,E_S=3-4h\,\,\,\,\,\,\big(\uparrow\uparrow\big)\,.
\label{d-7}
\ea
In the general case (\ref{d-5}) the minimization procedure can be simplified by making the change of variables
\be
\theta_A=\theta+\theta'\,\,\,{\rm and}\,\,\,\theta_B=\theta-\theta'\,,
\label{d-8}
\ee
leading eventually to
\be
E_S=3[(1+\Delta)x^2+(1-\Delta)y^2-1]-4hxy\equiv f(x,y)
\label{d-9}
\ee
with $x=\cos\theta$ and $y=\cos\theta'$ (notice the inversion symmetry $(x,y)\rightarrow(-x,-y)$ of (\ref{d-9})). If the Hessian ${\cal H}=36(1-\Delta^2)-16h^2$ is non-zero, then the only stationary point of $f$ is $x=y=0$ (meaning $\theta_A=0,\pi$ and $\theta_B=\pi,0$). For ${\cal H}>0$ this is a minimum point (since $f_{xx}>0$) and we have a N\'eel solid. In this case $h^2<(9/4)(1-\Delta^2)$, which in terms of $t$ and $\mu$ means
\be
\frac{3}{2}V-\frac{3}{2}\sqrt{V^2-4t^2}<\mu<\frac{3}{2}V+\frac{3}{2}\sqrt{V^2-4t^2}\,.
\label{d-10}
\ee
For ${\cal H}<0$, $x=y=0$ is an inflection point and the absolute minimum of $f$ then falls on the boundary of the domain, $[-1,1]^2$, precisely on $y=\pm 1$ (since in (\ref{d-9}) $y^2$ has a smaller coefficient than $x^2$). The minimum coordinates are simply calculated for $y=1$, or $\theta_A=\theta_B$ (a ground-state configuration that we can represent as $\nearrow\!\!\nearrow$ or $\searrow\!\!\searrow$). In this case $E_S=3(1+\Delta)\cos^2\theta-4h\cos\theta-3\Delta$ and, provided that $\big|\frac{2h}{3(1+\Delta)}\big|<1$, a minimum occurs for $\theta=\theta_m=\arccos\frac{2h}{3(1+\Delta)}$. This is also an absolute minimum and (since the spin component in the $x$ direction is non-zero) the system is condensed/superfluid. However, as $|h|$ increases for fixed $t$, $\cos\theta_m$ becomes eventually $\pm 1$; in terms of the original variables, this first happens at the lines $\mu=3V+3t$ and $\mu=-3t$. Beyond these lines, the system ceases to be superfluid and becomes insulating ($\theta_A=\theta_B=0$ or $\theta_A=\theta_B=\pi$).

In the superfluid phase, the grand potential (including the constant factor $C/V$ previously ignored) is
\be
E_S=V\big[3(1+\Delta)\cos^2\theta_m-4h\cos\theta_m-3\Delta\big]+3V-4\mu=-\frac{4(\mu+3t)^2}{3V+6t}\,,
\label{d-11}
\ee
the average occupancy is
\be
\rho_A=\frac{1}{2}+S_A^z=\frac{1}{2}(1+\cos\theta_m)=\frac{\mu+3t}{3V+6t}\,,
\label{d-12}
\ee
and the superfluid order parameter is
\be
\phi_A=S_A^x=\frac{1}{2}\sin\theta_m=\frac{\sqrt{(\mu+3t)(3V+3t-\mu)}}{3V+6t}\,.
\label{d-13}
\ee
In conclusion, all MF boundaries and characteristics of the QCT model perfectly match with those calculated in Ref.~\cite{Prestipino6} using the language of second-quantized operators.

\end{document}